\documentclass[preprint2]{aastex}
\def\kms{\rm{km \ s^{-1}}}
\def\etal{\rm{et al. }}
\def\deg{^\circ }
\def\solar{_\odot }
\clearpage

\begin{document}
  
\title{
Distances from Stellar Kinematics for Peculiar Virgo Cluster Spiral Galaxies
}
\author{Juan R. Cort\'es} 
\affil{Departamento de Astronom\'{\i}a, Universidad de Chile}
\affil{Casilla 36-D, Santiago, Chile}
\email{jcortes@das.uchile.cl}
\affil{National Astronomical Observatory of Japan}
\affil{2-21-1 Osawa, Mitaka, Tokyo, 181-8588}
\author{Jeffrey D. P. Kenney}
\affil{Department of Astronomy, Yale University}
\affil{P.0. Box 208101, New Haven, CT 06520-8101}
\email{kenney@astro.yale.edu}
\author{Eduardo Hardy\altaffilmark{1}}
\affil{National Radio Astronomy Observatory}
\affil{Casilla El Golf 16-10, Las Condes, Santiago, Chile}
\email{ehardy@nrao.edu }
\affil{Departamento de Astronom\'{\i}a, Universidad de Chile\altaffilmark{2}}
\affil{Casilla 36-D, Santiago, Chile}

\altaffiltext{1}{The National Radio Astronomy Observatory is a facility of the National
Science Foundation operated under cooperative agreement by Associated
Universities, Inc.}
\altaffiltext{2}{Adjoint Professor}
\received{someday} 

\shorttitle{Some title}
\shortauthors{Cort\'es, Kenney \& Hardy}

\begin{abstract}

We present distance estimates for eleven peculiar Virgo
cluster spiral galaxies based on measurements of the stellar kinematics of their
central 2 kpc.
Stellar circular velocities were obtained using two-integral dynamical models.
Distances were obtained by comparing, at each radius,
the stellar circular velocities with synthetic H$\alpha$ rotation curves
derived from NIR Tully-Fisher relations.
The results show that most of our galaxies are located within 4 Mpc of
the core of the cluster.
Three of these galaxies, previously classified as
``low rotator galaxies" or with ``Truncated/Compact" H$\alpha$ radial
distributions,
have stellar kinematics-based distances that are discrepant with HI-based
distances by at least 60\%,
and are likely to be located within the virial radius of the cluster.
These discrepancies appear due to very truncated
gas distributions plus non-circular gas motions or gas motions not in
the plane
of the stellar disk, perhaps as the result of gravitational interactions.
Our results show that
environmental effects can significantly reduce the measured HI linewidths
for some disturbed cluster galaxies, thus affecting
the accurate determination of distances based on gas kinematics methods.

\end{abstract}


\bigskip
\newpage

\section{Introduction}

For many years, the Tully-Fisher relation (Tully \& Fisher, 1977)
has been used to estimate
distances to the Virgo Cluster. These studies were based on optical or near
infrared photometry (Gavazzi \etal 1999), together with gas kinematics from
HI linewidths
(e.g., Yasuda \etal 1997) or H$\alpha$ rotation curves (Rubin \etal 1999).
These techniques have been useful in the determination of distances to
individual galaxies within the cluster, allowing attempts to determine
its 3-D structure (i.e. Yasuda \etal 1997, Gavazzi \etal 1999, and more
recently Solanes \etal 2002).

Recent results show that there is a population of galaxies in Virgo with HI
line--widths narrower
than those expected from the mean Tully-Fisher relation derived for the
cluster. We
refer to these as ``Tully-Fisher deviant" galaxies.
Either these galaxies are foreground to the Virgo cluster,
or they have peculiarities which make their observed gas line--widths
poor tracers of galaxy mass, and hence luminosity.

The Rubin \etal (1999) study of H$\alpha$ rotation curves of 89 Virgo disk
galaxies identified a population ( $\sim$ 6\% of total sample)
which appears to deviate
significantly from the Tully-Fisher relation, showing anomalously ``low
gas rotation" velocities at apparently every radius, which they called ``low rotators galaxies''.
Solanes \etal (2002) studied the 3-D structure of the cluster using
the Tully--Fisher relation based on HI linewidths and, most recently,
13 HI deficient spiral galaxies with apparently large 3-D barycentric
distances located far away in front and behind the core of Virgo
(Sanchis et al. 2004).
Seven galaxies were identified as being
in the foreground of the cluster core ($D <$ 13 Mpc),
supposedly in areas where the ICM
is too tenuous to be responsible for the stripping of the ISM gas from these
galaxies. Of these galaxies
only three had large projected distances ($\theta >$ 7$\deg$)
, whereas the rest had small projected
distance ($\theta <$ 4$\deg$). Some of the Solanes's
small HI line--width galaxies are also H$\alpha$ Rubin ``low rotator" galaxies.

In subsequent studies by the Sanchis \etal group
(Sanchis \etal 2002, Mamon \etal 2004, and Sanchis \etal 2004),
the possibility was analyzed that
these 13 HI-deficient objects had crossed the core of the cluster, losing
their gas content as a result of
ICM-ISM stripping. They concluded these galaxies
lie farther away than the maximum radius
to which these galaxies
can bounce out, making core-crossing improbable, and thus difficult for
ICM-ISM stripping to be the cause of the observed HI-deficiency. These authors
suggested alternative scenarios including the possibility of erroneous distance
estimations, HI-deficiency caused by tidal-interactions, heating of the gas by
mergers, and even questioned the claims for HI-deficiency.

The application of the techniques discussed above to HI deficient galaxies
suffers from the possibility that environmental effects can reduce the
spatial extent of
the atomic gas, cause non-circular motions or other disturbances,
thus reducing the measured
line-widths and therefore causing derived distances
to be underestimated
(Guhathakurta \etal 1988 and Teerikorpi \etal 1992). Indeed these small
line-widths could be caused either by truncation of the gas disks by ICM-ISM
stripping, or because the gas velocities are intrinsically small at every
radius (e.g. Rubin \etal 1999), which can happen if the gas motion is
non-equilibrium or non-circular,
or the gas geometry is not understood.

These conclusions raise obvious questions about these objects, such as:
Could these ``Tully-Fisher deviant'' galaxies be the result of an
underestimation of the distances produced by environmental effects,
leading to an artificial distortion of the true cluster shape?

We have carried out an extensive study of the kinematics and morphology of
13 peculiar Virgo Cluster
galaxies
(Cort\'es  \etal  2006; Cort\'es \etal 2008 in prep). 
Six galaxies within our sample, which includes four tentative
foreground galaxies, are objects with small HI line-widths and therefore
``Tully-Fisher deviant''. In order to answer these questions, in this paper
distances to
these galaxies are estimated through an approach that relies on stellar
kinematics\footnote{
Unfortunately, two of the original sample galaxies; NGC 4457 and NGC 4698, are unsuitable for
deriving their distances using
stellar kinematics since NGC 4457 is almost face--on, so its rotation curve is
uncertain, and NGC 4698
has complex kinematics (see Bertola \etal 1999; Falc\'on-Barroso \etal 2006
; Cort\'es \etal 2008 in prep) which makes its dynamical modelling
challenging, and escapes the scope of this work.} and is therefore less affected by environmental effects.

This paper is structured as follows. We describe in \S 2 the galaxy sample.
In \S 3,
the method for estimating distances via stellar kinematics is introduced.
The detailed application of this method to our data, the results on the distances to individual
galaxies of the sample, and the nature of
the ``Tully-Fisher deviant'' galaxies is studied in \S4.
A discussion about the
existence of the putative foreground group of galaxies in Virgo as well as of
the causes for these galaxies HI-deficiency is presented in \S 5.
Finally, suggested improvements to the technique,  
summary and concluding remarks are found in \S 8.

\section{The galaxy sample}

The sample consists of eleven
peculiar Virgo cluster spiral galaxies,
spanning a variety of optical
morphologies (Table $\ref{table1}$). Our sample includes more
early type than late type galaxies, since most of the strongly
disturbed cluster galaxies are early types.
Morphological selection was made using the R and H$\alpha$
atlas of Virgo cluster galaxies of
Koopmann \etal (2001), whereas the kinematical selection made use of
the published
H$\alpha$ kinematics on 89 Virgo cluster spirals by Rubin \etal (1999).

\begin{figure}
\centerline{
\plotone{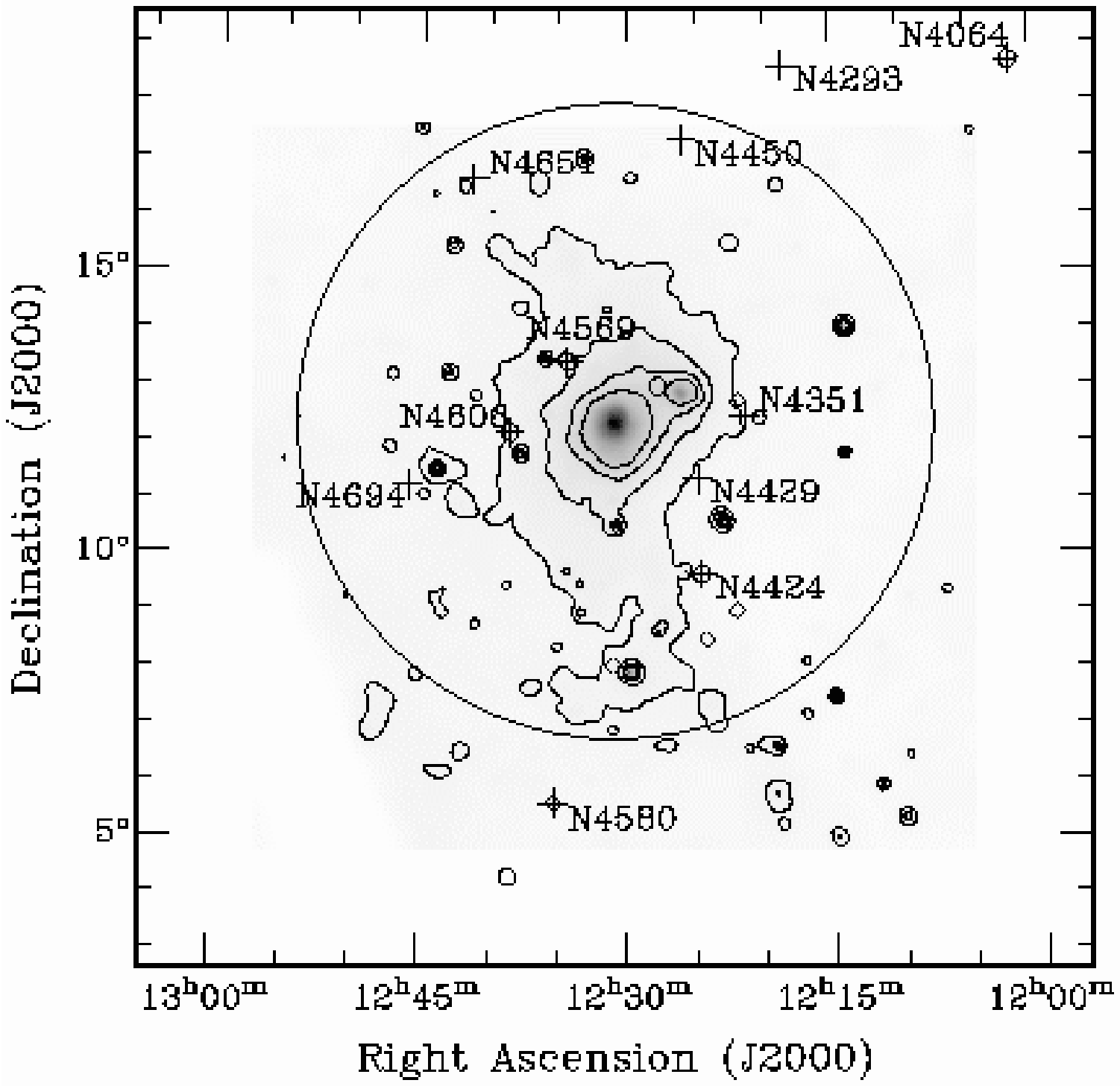}
}
\caption
{
Sample galaxies and their location in the Virgo cluster.
\rm
Contour map represent the ROSAT X-ray emission in the cluster (Bohringer \etal 1994).
Sample galaxies are represented by a black cross and its NGC name. Solanes's foreground galaxies
are represented by a circle, whereas background galaxies are represented by a
diamond. Four sample galaxies overlap with Solane's foreground galaxies, and one with
Solanes's background galaxies. Black circle represent the virial radius ($\sim$ 5.6$^{\circ}$)
around M 87.}
\label{virgosampletully}
\end{figure}

The Virgo cluster has about 110 spiral and lenticular galaxies brighter
than 0.1 $L^{*}_{\odot}$
(Bingelli \etal 1985, BST), so we have observed about 12\%
of the brightest Virgo cluster spiral galaxies. The
sample galaxies are spread around the cluster core, spanning projected distances
ranging from a few hundred kiloparsecs to 2.5 Mpc (Fig. $\ref{virgosampletully}$).

While the sample selection is not uniform, it is
designed to include bright Virgo spirals
whose peculiarities are most poorly understood, and to
include representatives of the different H$\alpha$ types identified
by Koopmann \& Kenney (2004).
In choosing sample galaxies within a given H$\alpha$ type,
we gave preference to those with kinematical peculiarities.

Following the use of the star formation classes introduced
by  Koopmann \& Kenney (2004),
we can group our sample galaxies as follow

\begin{itemize}

\item {\em Normal galaxies:} NGC 4651.

\item {\em Truncated/Normal galaxies:} NGC 4351, NGC 4569, NGC 4580,
and NGC 4694.

\item {\em Truncated/Compact galaxies:} NGC 4064, NGC 4424, and NGC 4606.

\item {\em Anemic or Truncated/Anemic galaxies:} NGC 4293, NGC 4429, and NGC 4450.

\end{itemize}

Although our main goal is to study galaxy morphological evolution in the
cluster,
we have at least six galaxies in the sample with small HI linewidths
(NGC 4064, NGC 4351, NGC 4424,
NGC 4580, NGC 4606, and NGC 4694), including four galaxies
that Sanchis \etal (2004) suggest may belong to a foreground group.

\section{Distance estimations by using stellar kinematics}

The determination of distances to HI-deficient galaxies using HI line-widths is vulnerable to the  reduction in the spatial extension
of the HI disks resulting from environmental effects, and by the possibility that gas velocities be intrinsically smaller at every radius.
On the other hand, the use of stellar kinematics for distance determination
(``stellar kinematics based distances", hereafter SKB distances)
has the advantage that it is in principle unaffected by physical processes other than gravitational processes,
whereas
as we have seen, the kinematics of the gaseous content can be affected by a number of non-gravitational physical
processes resulting from environmental interactions (e.g. ICM-ISM stripping).
However, stellar kinematics are far more difficult to measure and analyze than gas velocities,
on account of the faintness of the stellar component as well as the collisionless nature of the stars
which makes the pressure supported component important. The latter implies that rotation velocities
are no longer fully representative of the gradient of the potential.
Star velocities however, are as good tracers of the mass of the galaxy as gas velocities are,
provided one is able to properly handle their physical complexities.

We must therefore remember that; a) it is important to take into account the contribution of
the stellar velocity dispersion $\sigma$ to the real circular velocity, and b)
we have to restrict the Tully-Fisher analysis to a region within the galaxy where the surface brightness
is high enough to allow   reliable measurement of stellar velocities.

We have devised a distance estimation
method based on stellar kinematics rather than gas velocities. This method consists of the evaluation of
the circular velocity from the stellar kinematics through the use of two-integral dynamical models
(see Binney, Davies \& Illingworth 1990; van der Marel \etal 1990; Cinzano \& van der Marel 1994;
Magorrian \etal 1998; Cretton
\etal 1999) which are then used in a  Tully-Fisher relation. The method can be summarized as follow: 

1)  The stellar kinematics are modeled via two-integral dynamical models, to yield the true circular velocity $V_{c}$.

2) Because this circular velocity is valid only within the region that we have measured
reliably it cannot be used in a traditional Tully-Fisher
relation which requires sampling to large galactocentric distances. To resolve
this we construct synthetic ionized gas rotation
curves for galaxies spanning a range of absolute magnitudes, using near infrared
Tully-Fisher diagrams at each radius, as in Rubin \etal (1985). These synthetic rotation curves represent the typical rotation curves as a function of absolute magnitude for ``normal'' galaxies.

3) Finally, we compare these synthetic rotation curves with the circular
velocities derived in 1). The best match  gives the absolute
magnitude, and therefore the distance modulus.  

In the following sections we provide a more detailed explanation of
the estimation of the circular velocity and the construction of the
synthetic rotation curves.

\subsection{Estimation of the circular velocity from stellar rotation curves}

The Multi-Gaussian Expansion MGE formalism
(Emsellem \etal 1994; Cappellari 2002) provides one of the
simplest ways to build reliable two-integral dynamical models.
The MGE formalism allows us to parametrize the surface brightness of a galaxy as a sum of gaussians
with major axis widths $\sigma_{i}$, axial ratios $q'_{i}$, and
luminosities $L_{i}$,  
from which we can find a unique luminosity density $\nu$ (Cappellari 2002), assuming axisymmetry.
This approach offers advantages over 
the standard bulge + disk decomposition as a description of the light distribution.
It provides an analytical 
and efficient way to do the luminosity deprojection, and it allows
kinematical quantities such as the gravitational potential, velocity dispersion,
and circular velocity to be estimated with a simple integration, thus simplifying
the calculations.
These models can succesfully reproduce the observed stellar kinematics in (presumably)
axisymmetric galaxies.
However, we must keep in mind that issues such as dust obscuration,
triaxiality, kinematically distinct features in the stellar disks,
and the assumed lack of dark matter can easily make these models
discordant with the observed kinematics. The bias introduced by these
issues are discussed in \S 3.3.

To estimate the circular velocity, two-integral self-consistent models under the
MGE formalism were built for eleven galaxies of the sample.
These models assume axisymmetry and $\sigma_{R} = \sigma_{z}$.
They probably do not reproduce properly the kinematics, since the galaxies
can exhibit important anisotropies in their velocity ellipsoid,
but for the purposes of this work these models are acceptable since our
aim is to estimate the circular velocity rather than fully reproduce the
internal kinematics.
It is important to notice, that these models are purely stellar
and do not include the contribution
from dark matter. In most of our galaxies our measurements are
restricted to the inner parts
($\sim$ 20\% of the optical radius) making it impossible to
constraint properly the dark matter halo
parameters. Thus, these SKB distances must in all the cases
represent a lower limit to the distances.
However, precisely because we are sampling the central regions
where dark matter is the least important
its
effect on our distances is probably minor.    
The detailed modeling proceeded as follows:

1) The MGE formalism was used to find the luminosity density $\nu(R,z)$ from
R-band flux--calibrated images (Koopmann \etal 2001), assuming an oblate system. The dust lanes were
carefully masked to minimize the effects of dust obscuration. 
For a given surface brightness $\Sigma(x,y)$,
the formalism finds the best MGE model (Fig $\ref{fig4}$), which can be de-projected to yield a
unique luminosity density. The inclination was assumed to be the minimum required for an oblate distribution. In most cases this inclination was consistent with the inclination obtained by Koopmann \etal (2001).

2) If we consider the axisymmetric case, following Cappellari (2002),
the MGE mass density can be written as
\begin{equation}
\rho(R,z)= \Upsilon \sum_{i=1}^{N}\frac{L_{i}}{(\sigma_{i} \sqrt{2 \pi})^{3}q_{i}} \exp{\left[- \frac{1}{2 \sigma_{i}^{2}} \left(R^{2} + \frac{z^{2}}{q_{i}^{2}} \right)\right]},
\end{equation}
where, $\Upsilon$ is the stellar mass-to-light ratio, $N$ is the number
of adopted Gaussian components, each with
major axis  width $\sigma_{i}$, intrisic axial ratio $q_{i}$, 
and luminosity $L_{i}$.

Also, the MGE gravitational potential
can be written as
\begin{equation}
\Phi(R,z)=\sqrt{\frac{2}{\pi}} G\Upsilon \sum_{i=1}^{N} \frac{L_{i}}{\sigma_{i}} \int_{0}^{1} {\cal H}_{i}(T) dT,
\end{equation}
where, $G$ is the gravitional constant, and ${\cal H}_{i}$ is
an dimensionless function
given by
\begin{equation}
{\cal H}_{i}(T) = \frac{\exp{\left[-\frac{T^{2}}{2{\sigma}_{i}^{2}} \left(R^{2}+\frac{z^{2}}{1-\epsilon_{i}T^{2}} \right) \right]}}{\sqrt{1 - \epsilon_{i} T^{2}}},
\end{equation}
with $\epsilon_{i}=1-q_{i}^{2}$.

The circular velocity at each radius in the plane of the galaxies was
calculated directly from the potential.
The models have as input parameters the mass-to-light ratio
$\Upsilon$ and the distance $d$, but the luminosity scales with
the distance as $L \propto d^{2}$. This
lead us to conclude that the circular velocity
obeys
\begin{equation}
V_{c}^{2}(R)=R \frac{\partial \Phi}{\partial R}= \alpha {\cal F}(R).
\end{equation}
Here $\alpha \equiv \Upsilon d$, and ${\cal F}(R)$ is
\begin{equation}
{\cal F}(R) \equiv \sqrt{\frac{2}{\pi}} G\sum_{i=1}^{N} \left(\frac{L_{i}}{\sigma_{i}^{3}} \frac{R^{2}}{d} \right) \int_{0}^{1} {\cal H}_{i}(T) T^{2} dT,
\end{equation}
which represents the radial variation of the circular velocity,
and doesn't depend on the distance. Thus, the models can be
rescaled to any desired value of $\alpha$.

3) The  $\mu_{2} \equiv \sqrt{v_{LOS}^{2}+\sigma_{LOS}^{2}}$ moment was
calculated, for any point in the sky, following equations 61--63 from
Emsellem \etal (1994), after taking into account the typo error noticed by
Cappellari \etal (2006). To include the effects of seeing in the
computation of $\mu_{2}$, we weighted the data
when appropriate by the seeing convolved surface brightness. This moment has the advantage that
along the major axis it does not depend on the shape of the velocity
ellipsoid\footnote{Unfortunately,
along the minor axis it
depends critically on the ratio $\beta \equiv \sigma_{R}/\sigma_{z}$.}.

It can be shown that $\mu_{2}^{2}$ scales
with $\alpha$, which is the quantity we wish to determine. Thus, we
can choose a model with an initial guess for $\alpha$ (called
$\alpha_{0}$, see table $\ref{table5}$ for the
value of $\alpha_{0}$ used in each case ),
based on reasonable guesses for the galaxy distance
and $\Upsilon$, and rescale it to any mass-to-light ratio $\Upsilon$
and distance $d$ as  
\begin{equation}
\mu_{2}^{2}(\alpha)=\frac{\alpha}{\alpha_{0}} \mu_{2}^{2}(\alpha_{0}).
\end{equation}

4) Finally, the observed $\mu_{2}$ was calculated using the stellar
kinematics data from Cort\'es \etal 2008 (Fig. $\ref{kindata1}$).
The model-derived $\mu_{2}(\alpha_{0})$
moment and the observed $\mu_{2}$ were compared, and $\alpha$ was
calculated along the major axis of the galaxy as
\begin{equation}
\alpha = \alpha_{0} \left< \frac{{\mu_{2}^{2}}_{\rm obs}(R)}{\mu_{2}^{2}(\alpha_{0}, R)}\right>
\end{equation}
yielding the mean $\alpha$ factor  (e.g. Fig $\ref{fig6}$). Errors were estimated as
the standard deviation in $\alpha$. Therefore, the sought-after circular
velocity in equation (4) corresponds to that of the model which best
matches the observed $\mu_{2}$ giving the
best $\alpha$. It is important to notice that here we have derived
$\alpha$, but not $\Upsilon$ and $d$ independently. These should be
disentangled from $\alpha$.

\subsection{The Tully-Fisher relation for radii smaller than $R(V_{\rm max})$ }

So far we have derived the best circular velocity using two-integral dynamical
models. In what follows we focus on making use of these velocities to obtain the distances to our galaxies. 
In placing our circular velocities on a Tully-Fisher plot to derive
the distance and additionally disentangle $\Upsilon$ from the $\alpha$ factor, we must remember that we have 
measured velocities only until $R \sim$ 0.4 $R_{20.5}$ (where $R_{20.5}$
is the radius at which the galaxy has a surface brightness in H-band of
20.5 mag arcsec$^{-2}$). This makes it impossible, as discussed, to place
our velocities on the usual Tully-Fisher relation, usually determined at
$\sim$$R_{20.5}$, in order to obtain reliable distances. To overcome this problem,
we use the approach of building synthetic rotation curves based on
H$\alpha$ gas velocities as in Rubin \etal (1985). These synthetic rotation
curves represent the mean rotation curves for galaxies with different absolute magnitudes.

 The procedure to derive the synthetic rotation curves was the following:

 1) We constructed $H_{c}$-band infrared Pseudo Tully-Fisher (PTF)
relations for different normalized radii (0.05, 0.1, 0.2, 0.3, 0.6, 0.7, 0.8,
0.9 R$_{20.5}$), using the H$\alpha$ rotation curves for Virgo cluster galaxies
obtained by Rubin \etal 1999. (Fig. $\ref{ptf}$). The PTFs were determined
using the H-band magnitudes (Table $\ref{table4}$) derived by Gavazzi \etal
(1999), and Jarrett \etal (2003) for the galaxies of the Rubin's sample.
The advantages of using infrared magnitudes rather than $B$ magnitudes lie
in the reduction of the dependence of the Tully-Fisher relation on Hubble type,
in the reduction of the spread due to dust absorption and stellar population
variations, and in the improved ability to trace the luminous matter
(e.g.  Pierce \& Tully 1998, and Gavazzi \etal 1996). We excluded from the
T-F sample those galaxies that were identified as belonging to possible
background groups ($d > 23$ Mpc) by Solanes \etal (2002), whose could have
introduced undesirable systematic errors. We finally ended up with
34 galaxies, all with  H$\alpha$ kinematics and infrared photometry,
which we used to build the PTFs. We supposed that
all these galaxies were located at a mean distance of 16.8 Mpc, which
corresponds
to the distance to M 87 (Gavazzi \etal 1999; Nielsen \& Tsvetanov 2000; Tonry \etal 2001). The usual standard Tully-Fisher relation corresponds
to a PTF measured at $\sim$$R_{20.5}$. 

2) We fitted a straight line to each of the  above PTF relations corresponding
to a given radius. Then, for every magnitude, we derived an interpolated gas
velocity foe each radius from the  family of fitted straight lines (Fig
$\ref{ptf1}$, top panel). 

3) The final curves for each magnitude (and thus mass) were constructed using
the gas velocities obtained for each radius (Fig $\ref{ptf1},$ bottom panel).
These synthetic curves represent the expected shapes of the rotation curves
for galaxies with different luminosities (mass) belonging to the Virgo Cluster,
and obeying the Tully-Fisher relation. As we see in the graph, the synthetic
rotation curves do not overlap, depending only on the absolute magnitude.

The absolute magnitude corresponding to each synthetic curve was calculated by
assuming that each curve correspond to a galaxy located at a distance
$D_{V}=16.8$ Mpc.
Finally, the absolute magnitudes $M$ of our sample galaxies could be
found by locating the synthetic
rotation curve that best matches the stellar circular velocity curve for each sample galaxy. The latter
can be determined by
minimizing the $\chi^{2}$ (Figure $\ref{fig9}$) between the stellar
circular velocity and the synthetic rotation curves.
The distance modulus follows from the observed apparent magnitude $m$. Errors were  estimated using the variation
$\Delta \chi^{2}$, as 3--$\sigma$ errors ($\Delta \chi^{2}$=9).

Having derived the absolute magnitude by matching the best model
circular velocity and the synthetic rotation curves, the distance $d$ is
simply derived as
\begin{equation}
d=10^{\frac{m-M+5}{5}}.
\end{equation}
Moreover, after deriving $d$ we can obtain the mass-to-light ratio
as $\Upsilon = \alpha/d$.

The comparison between gas-based synthetic rotation curves and
stellar circular velocities is justified,
if we exclude the galaxies with peculiar rotation curves such as the low rotators,
since gas-based synthetic rotation curves represent the typical rotation curve, and thus the gradient
of the potential.
It is
important to notice however that discrepancies between gas-based synthetic rotation curves, and stellar
circular velocities could be found in the inner parts of the galaxies ($r \leq$ 0.1 $R_{20.5}$), where the
gas velocity dispersion or non-circular gas motions could be important enough to affect the determination of the gas-based
synthetic rotation curves, and peculiar stellar structures as circumnuclear disk can be found. Thus we
restrict our comparision between stellar circular velocity and gas-based synthetic rotation curves to
radii greater than 0.1 $R_{20.5}$.

\subsection{Biases in the stellar kinematics-based distance estimations}

Our SKB distances are not free of systematic effects. We recall that these effects
are:

1)  The presence of dust
in our galaxies can reduce the amount of luminous
mass that contributes to the potential, so we could underestimate the
expected circular velocity. This problem arises preferently in
galaxies with prominent dust lanes. 

2) Triaxiality and non-axisymmetric structures are ignored in the pursuit of  oblate models. This in turn introduces either an overestimate or underestimate of the mass
by altering the amplitude and shape of the
rotation curve. Galaxies with strong bars could be seriously affected.
Moreover, triaxial features are very difficult to model and their inclusion
is beyond the scope of this work.

3) Kinematically distinct components are also important. Some galaxies
have circumnuclear stellar disks (e.g. NGC 4429) or counter-rotating disks
which can be easily detected in the velocity fields but
are hard to detect photometrically. The latter makes the inclusion of such
components in the models difficult making our models discordant with the
observations in regions dominated by such features. Thus, we must exclude in
our comparison the regions where these features are important.

4)  The dark matter halo is an important component of the potential
of the galaxy especially in the outer parts.
The dark matter
contribution can be
represented by a logarithmic potential,
\begin{equation}
\Phi_{DM}=\frac{v_{0}^{2}}{2} \ln (r^{2} + r_{0}^{2}),
\end{equation}
which is the simplest potential producing a flat circular velocity $v_{0}$ at larger radii ($r >> r_{0}$).
In figure $\ref{figdm}$, we show the effect of the dark matter halo in two sample
galaxies; NGC 4450 and NGC 4569. The parameters $r_{0}$ and $v_{0}$ were chosen in
order to match the observed H$\alpha$ rotation curve from Rubin \etal (1999). This figure shows that
the dark matter halo is indeed important in the outer parts, but our observations are limited to
the inner 30", so we are unable to constrain the dark matter halo
parameters properly, implying that we are underestimating the circular velocity. 

Most of these problems imply an underestimation of the circular velocity. For example, dust and dark matter have the similar effect of hiding the amount of mass that contributes to the potential. This translates to an underestimate of the luminosity and therefore the model masses. Thus, we should consider these SKB distances as lower limits to the real distance;
galaxies will tend to be somewhat further away than estimated here.

\section{Distances to the peculiar Virgo cluster galaxies}

\subsection{The actual fittings}

\begin{figure}
\epsscale{0.8}
\plotone{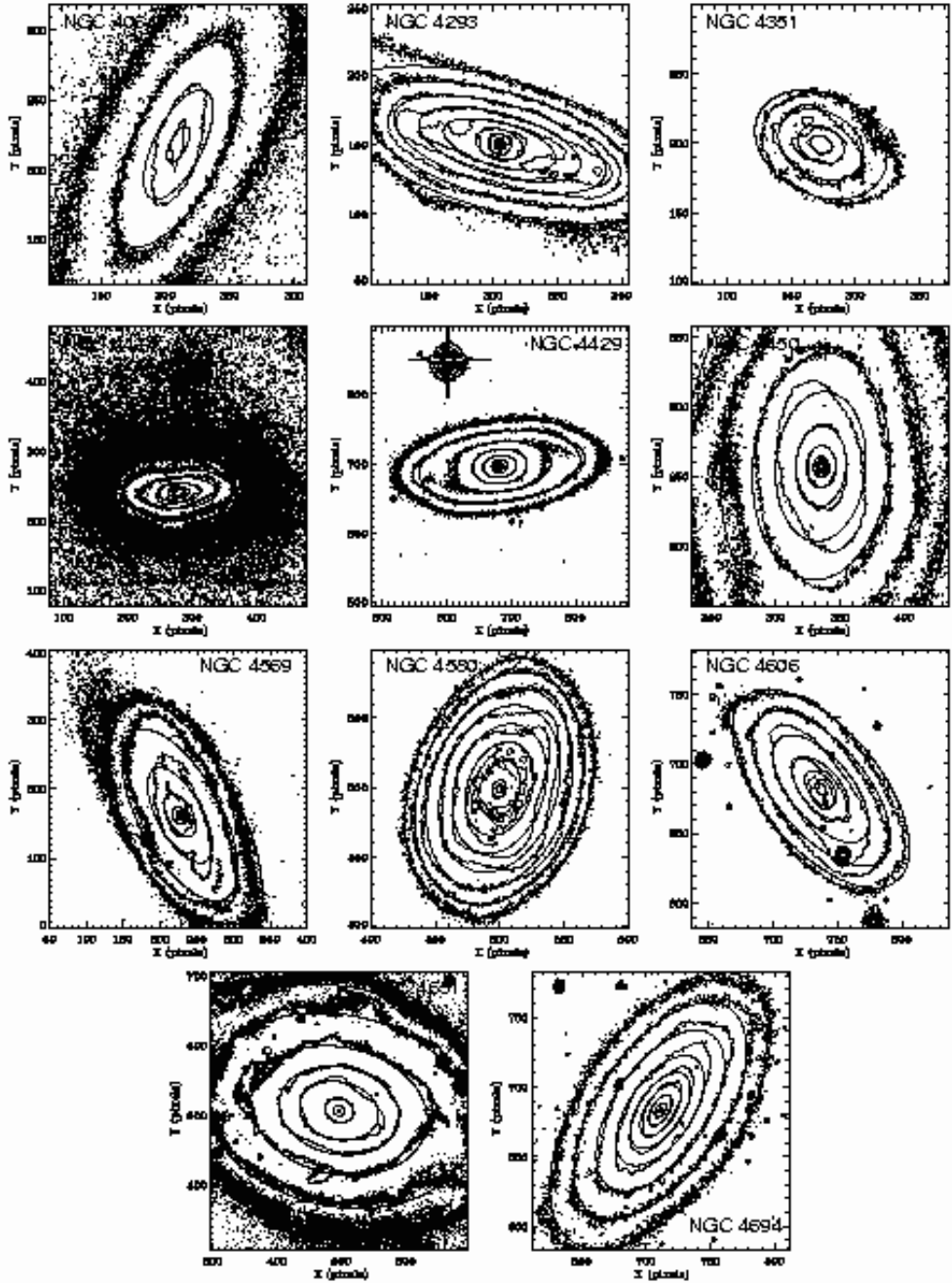}
\caption{MGE fits to sample galaxy R-band images.
MGE model contours are overplotted on R-band contours.}
\label{fig4}
 \end{figure}

\begin{figure}
\figurenum{3}
\includegraphics[width=1.0\textwidth, clip]{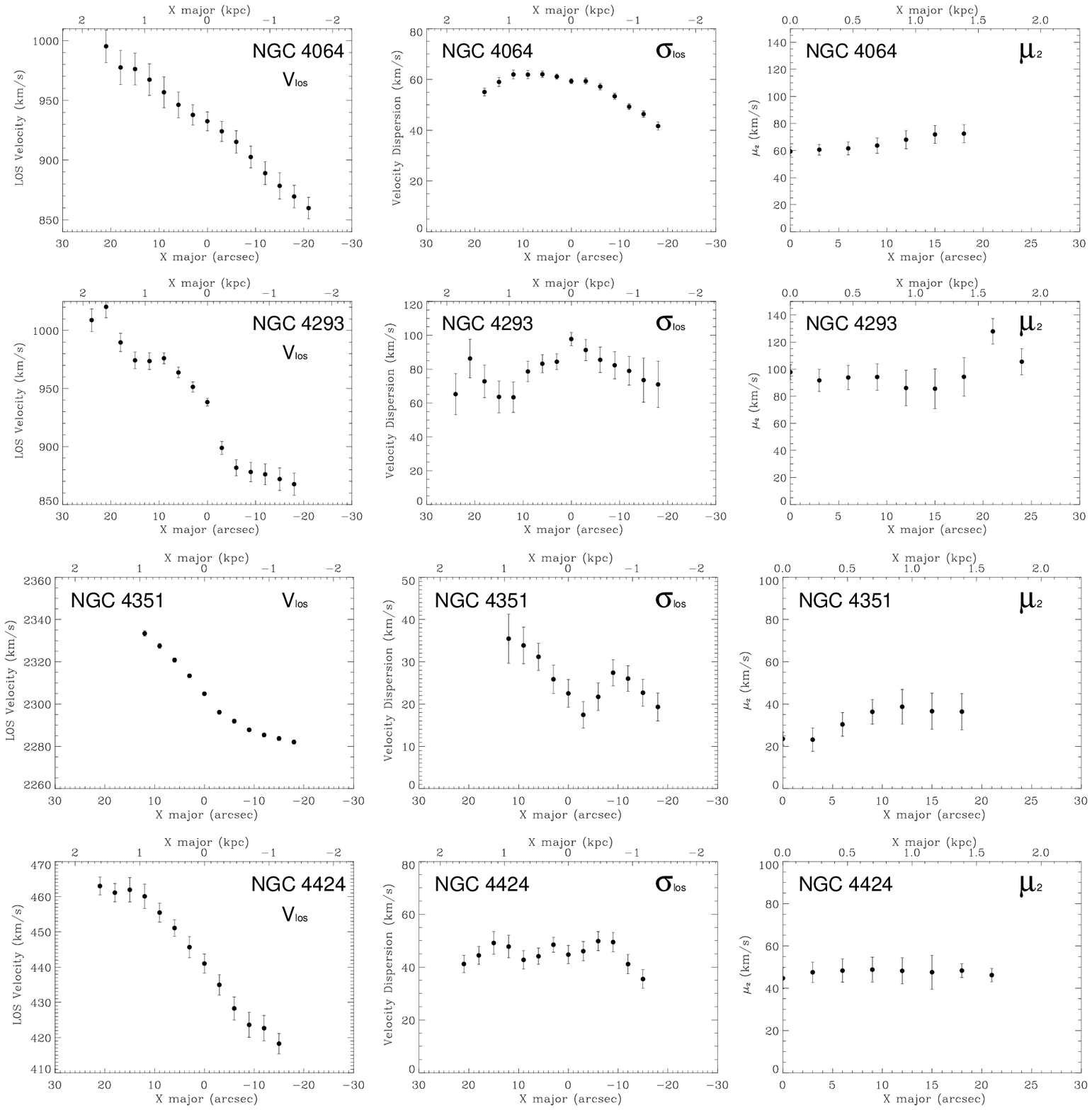}
\caption{Stellar kinematics along the major axes of the sample galaxies. 
From left to right: the stellar velocity, the velocity dispersion, and
the  $\mu_{2}$ moment.
}\label{kindata1}
\end{figure}

Now that we have discussed in detail the methodology, we focus on the actual
fittings and their errors. The MGE fits to the light distribution
are presented in Fig. $\ref{fig4}$.
We see that most galaxies are reasonably well fit by the MGE algorithm.
Differences in surface brightness between the galaxy images and the MGE model
are less than 10\%, which translates to velocity
uncertainties of the order of 5-10\%. Galaxies such as NGC 4064 and
NGC 4450 present bigger uncertainties resulting from the existence of a
central bar that obviously cannot be accounted by an axisymmetric
decomposition.

\begin{figure}
\figurenum{4}
\plotone{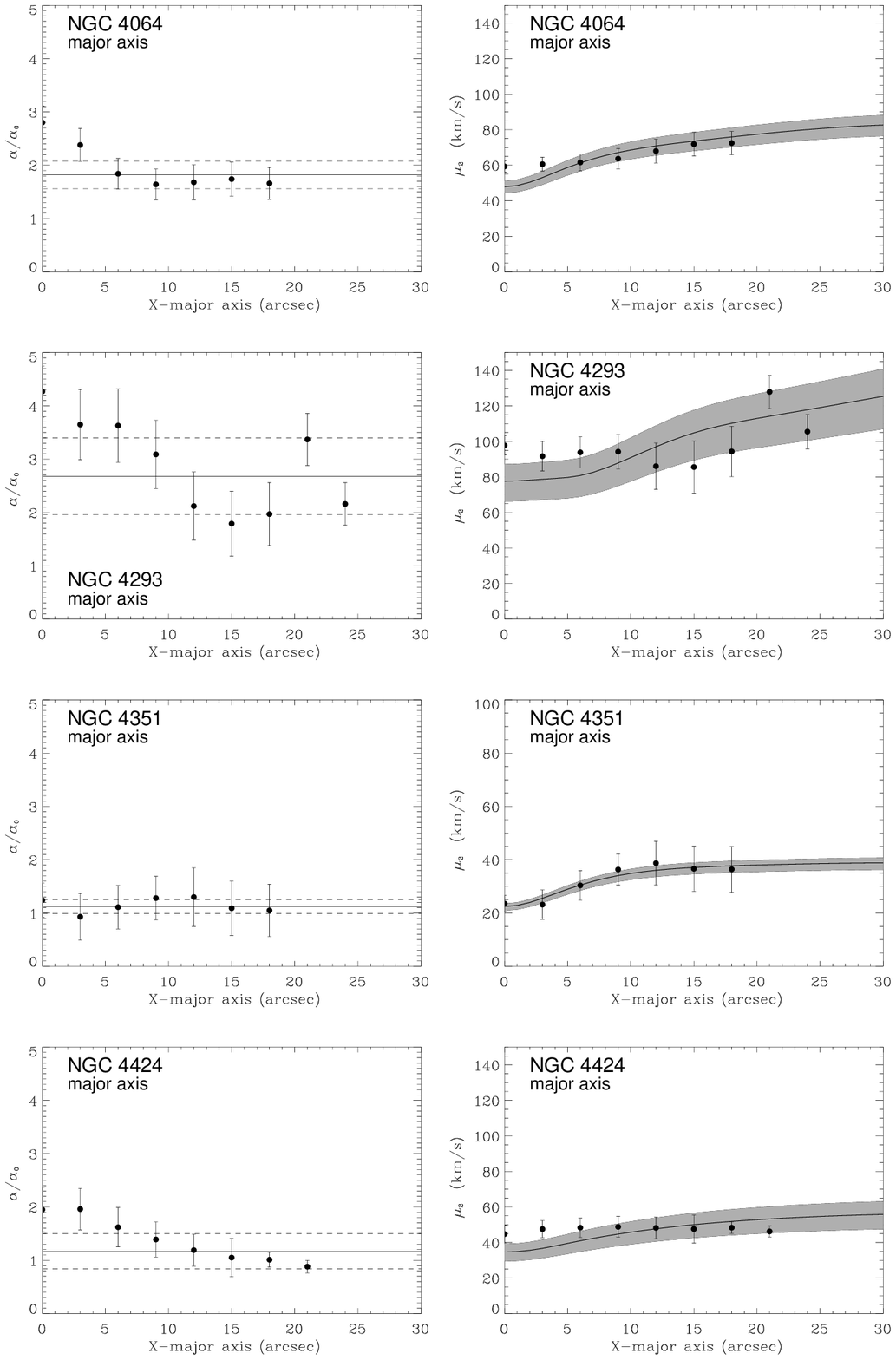}
\caption{$\alpha/\alpha_{0}$ ratios (left panels) and $\mu_{2}$ moments (right panels) along the major axes.
The best fit model is represented by thick solid lines, and the model uncertainties 
are indicated by dashed lines (left) or shaded areas (right). 
}\label{fig6}
 \end{figure}

\begin{figure}
\figurenum{5}
\centerline{
\plotone{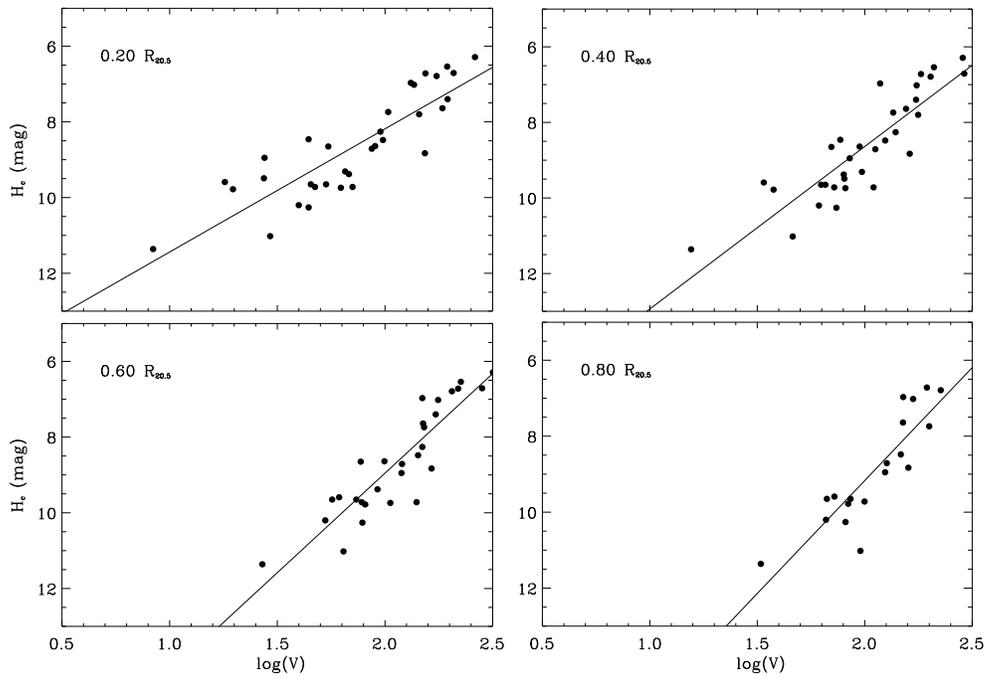}
}
\caption{Pseudo Tully-Fisher relations for different radii. Solid circles represent Rubin's H$\alpha$ velocities for Rubin's galaxy sample.
The lines represents linear fits to the Pseudo Tully-Fisher relations.}
\label{ptf}
\end{figure}

\begin{figure}
\figurenum{6}
\centerline{
\plotone{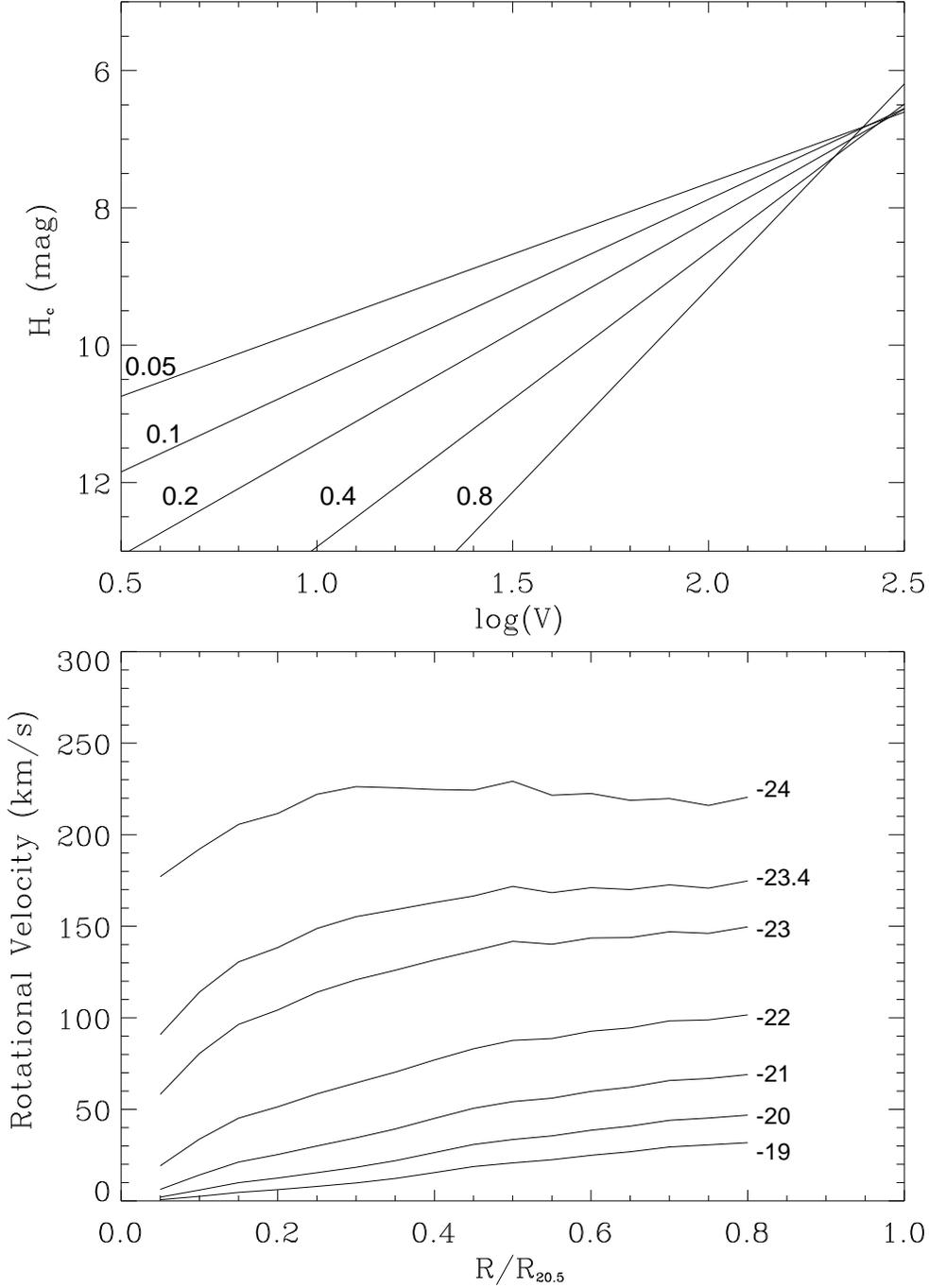}
}
\caption{
Pseudo Tully-Fisher relations and synthetic rotation curves.
{\em Top Panel:} Lines fitted to the Pseudo Tully-Fisher relation for different radii
(0.05, 0.1, 0.2, 0.4, 0.8 $R_{20.5}$). From the interception between those lines and
the line for constant magnitude, we find the corresponding rotational velocity for each radius.
{\em Bottom Panel:} Synthetic rotation curves derived from Pseudo Tully-Fisher relations labelled with corresponding values of absolute magnitudes $M_{H}$.}
\label{ptf1}
\end{figure}

\begin{figure}
\figurenum{7}
\centerline{
\includegraphics[width=0.7\textwidth,clip]{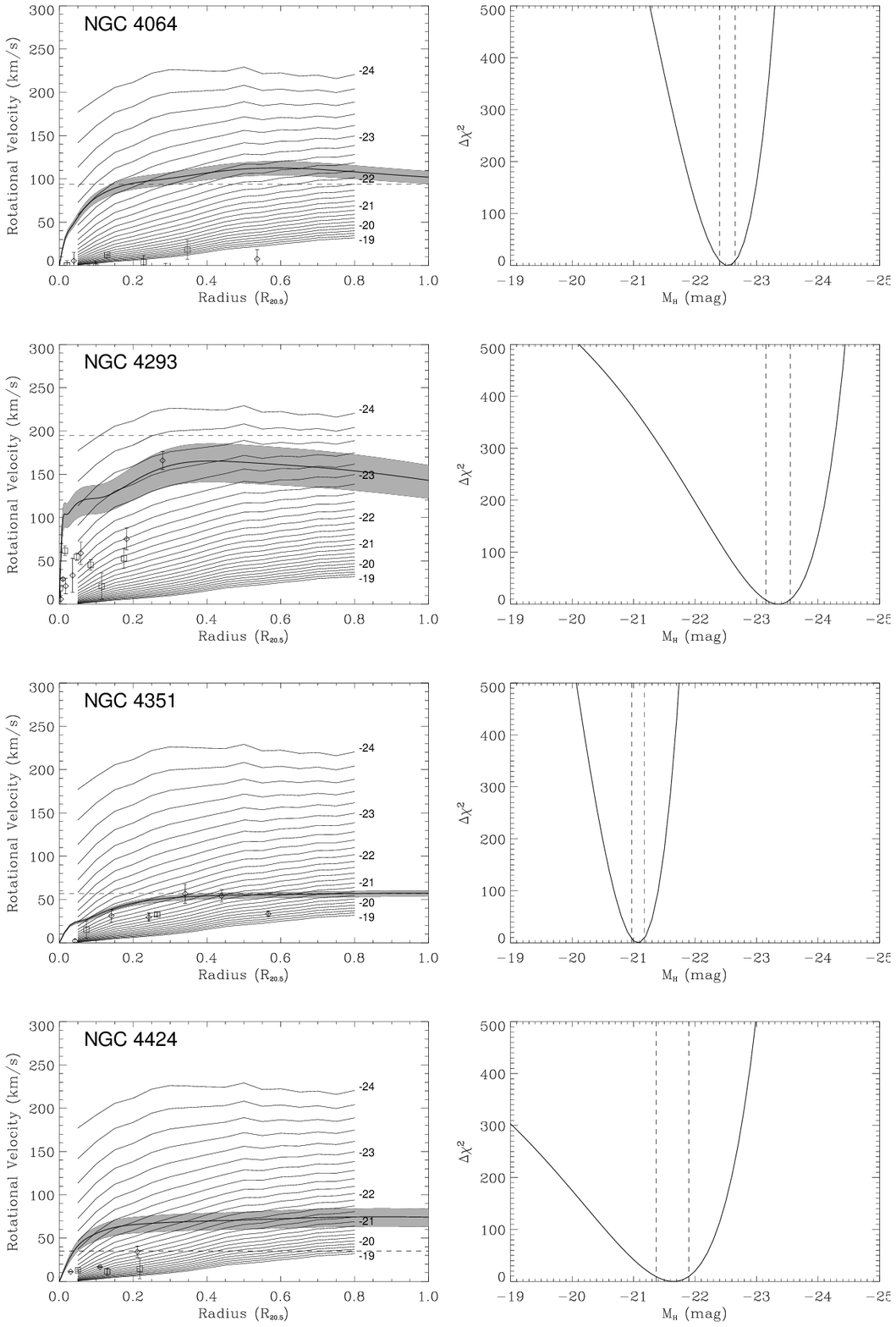}
}
\caption{MGE stellar circular velocities and synthetic rotation curves.
{\em Left panels;} stellar circular velocities.
Circular velocities (thick solid line with shaded area representing uncertainty) roughly match with one synthetic rotation curve (solid lines).
H$\alpha$ gas rotation curves are represented by open symbols. HI maximum velocities (Solanes
\etal 2002) are represented by the dashed line.
The numbers to the right of the synthetic rotation curves represent the absolute $H$ magnitudes for a 
galaxy having such a rotation curve at the distance of the Virgo cluster.
{\em Right panels;} $\Delta \chi^{2}$ between the MGE circular velocity and
synthetic rotation curves, dashed lines represent the 3--$\sigma$ confidence
level.
}\label{fig9}
\end{figure}

The $\alpha/\alpha_{0}$ ratio salong the major axis are displayed in Fig $\ref{fig6}$. In most
galaxies, the values are roughly constant along the major axis outside
the central 5", except in NGC 4293, NGC 4424, NGC 4429, and NGC 4450. In NGC
4429 there is a prominent circumnuclear disk which cannot
be reproduced with these simple models. The best $\alpha/\alpha_{0}$ ratios,
therefore $\alpha$, were
determined (Table $\ref{table5}$) as the average along the major axis outside the central 5", 
except for NGC 4429 where we use distances greater
than 12" in order to exclude the circumnuclear disk. The average $\alpha/\alpha_{0}$ ratio is displayed as a solid line in figure $\ref{fig6}$ (left panels),
while the errors appear displayed as dashed lines.
The comparisons between the observed and best model $\mu_{2}$ moments along the
major axis for the galaxies are displayed in Fig $\ref{fig6}$ (right panels)
On these plots,
the solid line represents the best model defined by the mean $\alpha$
factor, the shaded region represents the range of uncertainty given
by the standard deviation
in $\alpha$. Along
the major axis, the model and the observed $\mu_{2}$ in most cases
agree reasonably. 

To compare the synthetic rotation curves with the circular velocity curves,
the latter were rescaled in radius in terms of $R_{20.5}$ in the H-band  (Gavazzi \etal 1999). The exceptions were
 NGC 4064 and NGC 4424 whose values were determined from H-band surface
brightness profiles derived by Jarrett \etal (2003), since Gavazzi \etal (1999) do not
have measurements for these galaxies.  This is shown in Fig. $\ref{fig9}$. The model-based circular velocities
extend further than the stellar kinematic data. This is justified since the circular
velocities are derived from self-consistent models 
based on optical images which extend into the outer galaxies.

The comparison between the stellar circular velocities and the
synthetic rotation curves were performed by monitoring the $\chi^{2}$
between the synthetic rotation curves and the stellar circular velocities,
within the range of 0.1 -- 0.8 $R_{20.5}$. This range was chosen in order
to avoid 
large variations in velocity existing within the inner 0.05 $R_{20.5}$ (e.g. 
NGC 4429, NGC 4450, and NGC 4569), and to minimize the
effect of not including a dark matter halo, which can be significant
if we compare only the outer parts (i.e., using a range of radii of
0.6 -- 0.8 $R_{20.5}$).
Comparison between the present circular
velocities with those derived by Rubin in H$\alpha$ show that in most cases
Rubin's values are of the order of or lower than our circular velocities.
The exceptions are NGC 4450, and NGC 4569. In both cases, this
implies that
the galaxies are more massive than  predicted by our models. In the case of
NGC 4450, the MGE fit
(Fig. $\ref{fig4})$ exhibits a big discrepancy in P.A on one of the isophotes,
indicating the presence of an important non-axisymmetric structure which might
account for the missing mass. In NGC 4569, dust lanes are prominent until at
least 5 kpc through the disk, which prevented us from including all the
luminous mass.

\subsection{The SKB distances}
Once the absolute magnitudes M$_{H}$ for each galaxy is determined as is described in
section \S 3.2, the line-of-sight distance can be found from the distance
modulus. $\Upsilon$ can also be derived from $d$ and $\alpha \ $(Table
$\ref{table6}$). To test whether the results on $\Upsilon$ are physically meaningful,
we plot in Fig $\ref{virial}$ the relation between the R-band luminosity, and
the R-band mass-to-light ratio. The best fit straight line in $\log L$, $\log \Upsilon$ space is
\begin{equation}
\log(\Upsilon/\Upsilon_{\odot})=-3.2 \pm 0.6 +(0.35 \pm 0.05) \log(L/L_{\odot})
\end{equation}
This fit is roughly consistent within the errors with the fundamental plane
correlation
$\Upsilon \propto L^{0.2}$ predicted from the virial theorem (e.g., Faber \etal 1987; Bender, Burstein, \& Faber 1992). This consistency lends credibility
to the  modelling presented here. The observed variations in $\Upsilon$ are mostly due to the variation in $L$. However, there are galaxies that present very low $\Upsilon$.
For example, NGC 4694 has a value of $\Upsilon$ of 1.1 $\pm$ 0.3 $\Upsilon\solar$, which is a factor $\sim$ 2 lower than the expected. This low value and its
$B-V$ color ($\sim$ 0.62), suggest a recent starburst.

Figure $\ref{fig11}$ which displays the galaxy distance distribution shows that
our sample galaxies are located between 10 to 27 Mpc from us,
with the majority of them between 13 and 20 Mpc. Two galaxies,
NGC 4450 and NGC 4569, have model locations in front of the Virgo main core at about
10--12 Mpc from us, although the models underestimate the mass for these
galaxies, so it is likely that these galaxies
are not foreground (see notes in table $\ref{table6}$). On the other hand,
NGC 4651 is located at 27 Mpc and is clearly
a background galaxy. 

It is interesting to compare our stellar kinematics-based distance estimates 
with Solanes's HI-based distance estimates.
In figure $\ref{statdiff}$, we show the correlation between Solanes's
distances and the ratio between the SKB distances and Solanes's, with 3--$\sigma$ error bars.
>From inspection of  this figure, we conclude that 
NGC 4293, NGC 4450, NGC 4569, NGC 4580, and NGC 4651 have SKB distances
that agree within the errors with those of Solanes. But others such as
NGC 4064,
NGC 4351, NGC 4424, and NGC 4606 have significant
discrepancies worthy of further analysis.

\subsubsection{The cases of NGC 4064, NGC 4424 and NGC 4694: Rubin's low rotators}
     
NGC 4064 and NGC 4424 are particularly interesting because both galaxies have
been regarded as possible foreground galaxies.
Moreover, both galaxies, together with NGC 4694, belong to the
class of Rubin's ``low rotator" galaxies, although
for the case of NGC 4064 Rubin's H$\alpha$ velocities are much smaller
than our ionized gas velocity measurements (Cort\'es \etal 2006). Solanes's distance estimations
indicate a distance of
9.8 $\pm$ 0.4  Mpc for NGC 4064, and 4.1 $\pm$ 0.2 Mpc for
NGC 4424\footnote{ 1$\sigma$ errors} in stark contrast with our
estimations of 18.0 $\pm_{0.8}^{1.3}$ Mpc, and 15.2$\pm 1.9$ Mpc respectively.
The discrepancies are of a factor 1.8 for NGC 4064 and 3.7 for NGC 4424,
which are statistically significant.
In the case of the other ``low rotator", NGC 4694, we determine a distance of
13.4 $\pm_{1.0}^{1.3}$ Mpc. Clearly, NGC 4064 and NGC 4424 belong to
the Virgo cluster. NGC 4694 might be  affected by the dark matter bias in our
models, so its distance should be considered a lower limit. All these galaxies
present truncated H$\alpha$ disks (Koopmann \& Kenney 2004), and their gas
kinematics are peculiar (see Cort\'es \etal 2006), indicating that they are
heavily disturbed by the environment.
They all show evidence for strong gravitational interactions or mergers,
which may account for their low gas velocities.
This is cautionary of the risk of
using gas velocities as distance estimators for HI deficient and/or disturbed
galaxies. We conclude that the ``low rotators" found by Rubin do not exhibit
a low gas rotation velocity because they are intrinsically faint and thus
closer to us, but rather because they have been heavily affected by
gravitational interactions perhaps in addition to ram pressure stripping.

\subsubsection{NGC 4351 and NGC 4606}
Consider now the cases of NGC 4351, and NGC 4606. Both galaxies
belong to Rubin's sample but they were not classified as ``low rotators".
Solanes's estimation gives
distances of 11.2 $\pm$ 1.3 Mpc, and 12.7 $\pm$
0.3 Mpc respectively which again differ from our results of
20.3 $\pm_{1.2}^{0.8}$ Mpc, and 19.9 $\pm_{0.5}^{1.0}$ Mpc.
In NGC 4351 the discrepancy, could be due to truncation of the HI disk via 
ICM-ISM stripping, a very likely effect on this galaxy considering its high
relative velocity with respect to the cluster mean
($\sim$ 1200 $\kms$). NGC 4606 has a star formation class ``Truncated/Compact",
same as NGC 4064, and NGC 4424, and as these galaxies it has features that
suggest gravitational interactions (Cort\'es \etal 2006), so again its
small line-width could be due to the action of environmental effects.

\subsubsection{NGC 4569}

This galaxy has a SKB estimated distance of
9.9 $\pm 0.2$ Mpc, still in front of Virgo. 
However, the SKB distance for NGC~4569 may be an underestimate,
because of its strong dust lanes and  possibly large influence of its 
dark matter halo (Fig $\ref{figdm}$). If we include
the dark matter halo in order to match the Rubin's H$\alpha$ rotation curve we
find a barycentric distance of 3.8 Mpc, locating this galaxy inside the
maximum rebound radius. The SKB distance is slightly bigger than
Solanes's HI-based distances, although there are different HI linewidths
reported in the literature for NGC 4569, and one gets a different
distance depending on which linewidths one uses. 
The HI linewidth in figure $\ref{fig9}$ comes from the Arecibo General
Catalogue, a private database mantained by Riccardo Giovanelli and Martha P.
Haynes, partially published by Solanes \etal (2002). This is not necessarly
consistent with the homogenized HI-distance of Solanes \etal (2002), which is
an {\em average} between different Tully-Fisher (TF) catalogues. If fact, the
HI-distance in the TF catalogues used by Solanes \etal can be as low as
7.9 Mpc (Kraan-Korteweg \etal 1988) and as high as 16.1 Mpc (Ekholm \etal 2000).

\subsection{SKB distances and star formation classes}

As it was shown in \S 4.2, in the cases of NGC 4293, NGC 4450, NGC 4569, NGC 4580,
and NGC 4651, HI-based distances agree with our SKB distances. These galaxies
have star formation classes (Koopmann \& Kenney 2004) ranging from
``Truncated/Anemic" for NGC 4293, and NGC 4450, to ``Truncated/Normal" for
NGC 4569 and NGC 4580, and to ``Normal" in the case of NGC 4651.
In contrast for the ``Truncated/Compact" galaxies the SKB distances
don't agree with HI-based distances. This suggests that ``Truncated/Compact" galaxies are
not only very truncated in their gas content, but that they also have non-circular
gas kinematics or gas motions which are not in the plane of the stellar disk.
 
Here we comment on the possible existence of several galaxies in the foreground 
of Virgo at $\sim$10 Mpc (Solanes \etal 2002), which were tentatively identified as a group by
Sanchis \etal (2002). Four of our sample galaxies are considered foreground
by Solanes. All of them except for NGC 4569 are ``Truncated/Compact"
(i.e., NGC 4064, NGC 4424, and NGC 4606) with SKB distances in the range 16--20
Mpc. NGC 4569 has a stellar-kinematics estimated distance of
9.9 $\pm 0.2$ Mpc, still in front of Virgo.
However, this galaxy shows strong evidence for ongoing ICM-ISM stripping (Vollmer \etal 2004;
Kenney \etal 2008 in prep), and its angular distance
(1.7$\deg$) suggests that it lies within a region where the ICM medium is
dense enough for efficient stripping of its gas. It is important to recall that
our SKB distances should be considered as a lower limit due to the biases.
We should expect that the real distances are larger than our estimates, so
the discrepancies between Solanes' and our determinations should
increase if we correct for the biases.  

We conclude, on the basis of our overlapping sample (we have analyzed
60\% (4/7) of their foreground candidates), that there is no
compelling evidence for the existence of multiple foreground galaxies at D$\simeq$10 Mpc,
as suggested by Solanes \etal (2002).

\section{Barycentric distances and the origin of HI-deficient galaxies in the outskirts of the cluster}

In order to determine how far these galaxies are with respect to the core of
the Virgo Cluster, we calculated their barycentric distances $r$ to M87,
assumed to be at a distance of
$D_{V}$=16.8 Mpc (Gavazzi \etal 1999; Nielsen \& Tsvetanov 2000; Tonry \etal
2001). These distances can be derived from
\begin{equation}
r^{2}=D_{V}^{2}+d^{2}-2 D_{V}d \cos{\theta}
\end{equation}
where $d$ is the line-of-sight distance of the galaxies, and $\theta$ the
angular distance to M87. The results can be found in Table $\ref{table6}$ and
are shown in Fig $\ref{fig11}$. All galaxies are plotted on a P-V diagram.
The solid line represents the ``virial radius'' ($r_{\rm V} = $ 1.65 $h_{2/3}^{-1}$ Mpc, where $h_{2/3} = H_{0}$/(66.7 km s$^{-1}$ Mpc$^{-1}$)) derived by Mamon \etal (2004),
and the dashed line represents the largest``maximum rebound radius'' ($r_{reb} = $ 1--2.5 $r_{V}$)
of all the sample galaxies.
The latter corresponds to the maximum radius to which particles that cross the
core can bounce to. In their calculations Mamon \etal (2004) considered only
one single halo, the M 87 subcluster, which is the most massive substructure
in Virgo.

We see in Fig $\ref{fig11}$ that most of our galaxies are within the largest maximum
rebound radius, and such galaxies could have crossed the
core of the cluster and suffered ICM-ISM stripping,
which reducing the size of their gas disks.
While many of the HI-deficient spirals in the outer cluster probably lost most of their gas
during a core passage, others have apparently experienced ICM-ISM stripping in the cluster outskirts.
The Virgo spiral NGC 4522 is a clear example of outer cluster ICM-ISM stripping 
(Kenney \etal 2004, Crowl \& Kenney 2006), which is thought to be caused
by ram pressure which is enhanced due to an ICM which is dynamic and lumpy rather than static and smooth.

Others likely owe their HI-deficiency
to gravitational interactions, such as NGC 4064 (Cort\'es \etal 2006) and NGC 4293,
which exhibit kinematical peculiarities and disturbed dust distributions,
both signatures of gravitational interactions.

\begin{figure}
\figurenum{8}
\centerline{
\plotone{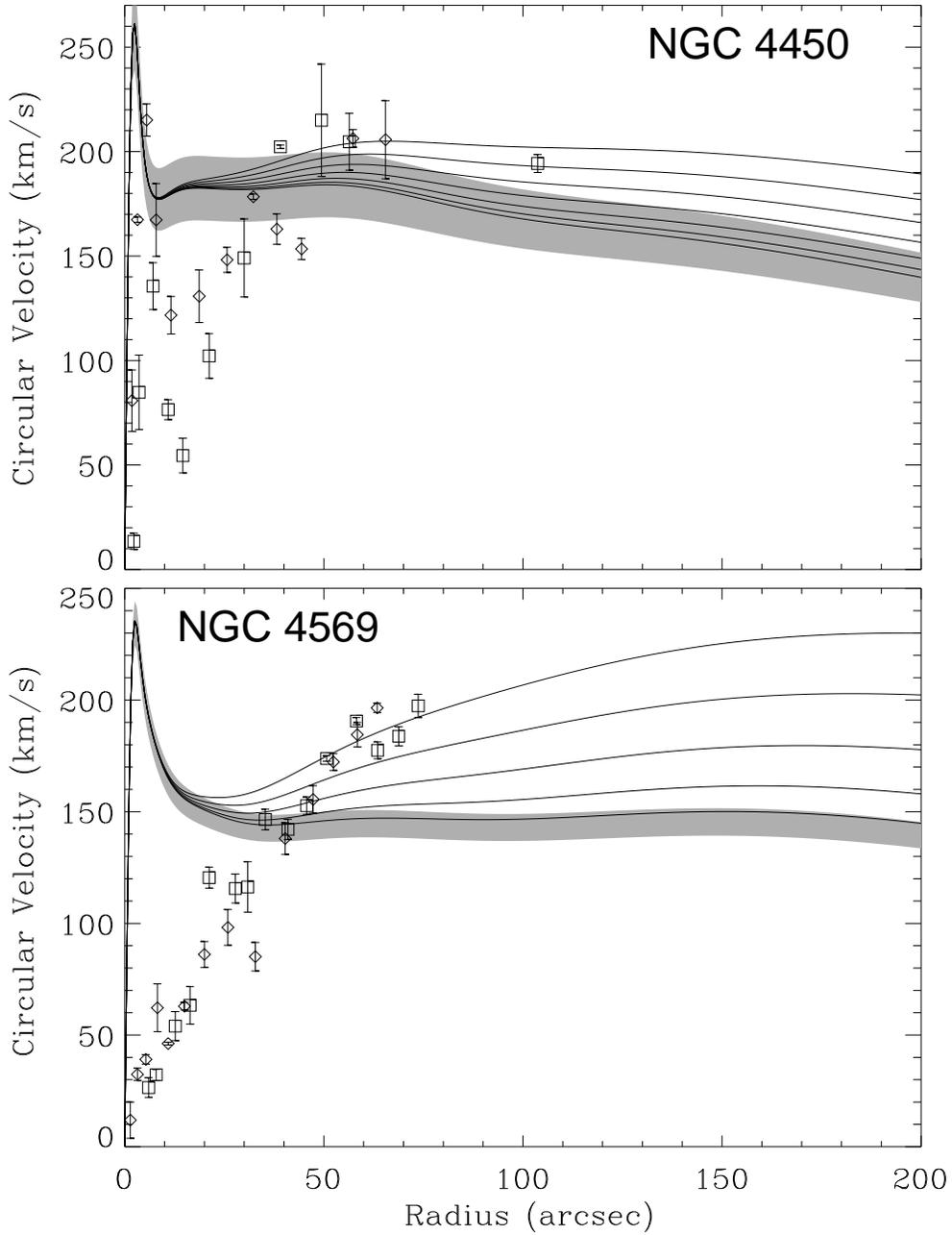}
}

\caption{Effect of the dark matter halo in two sample galaxies. Solid lines
represent the circular velocities for different dark matter halo parameters.
Open symbols represent the Rubin etal (1999) H$\alpha$ rotation velocities, and shaded areas represent the uncertainties in the pure stellar circular velocities. Our stellar
kinematics data is limited to the inner 30", so we cannot constrain the
dark matter halo parameters without using the ionized gas rotation curves, but even including them the dark matter halo parameters are too uncertain.}
\label{figdm}
\end{figure}

\begin{figure}
\figurenum{9}
\centerline{
\plotone{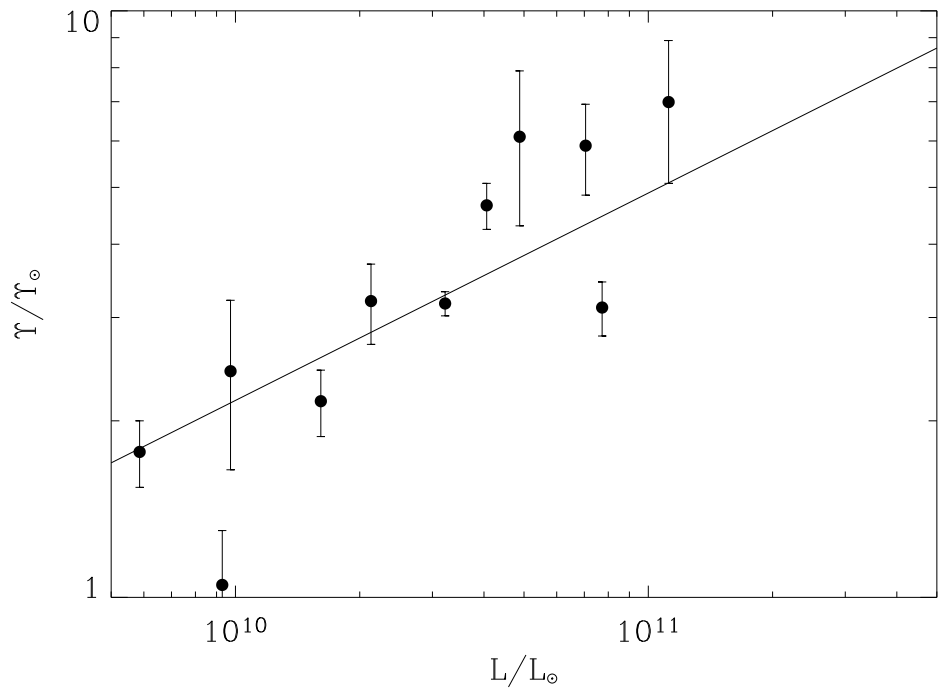}
}

\caption{Correlation
between the R-band stellar mass-to-light ratio $\Upsilon$ and the
luminosity $L$.
Solid line represents the fit of a straight line, which roughly
follows the fundamental plane relation $\Upsilon \propto L^{0.2}$.}
\label{virial}
\end{figure}

\begin{figure}
\figurenum{10}
\centerline{
\includegraphics[width=1.0\textwidth,clip]{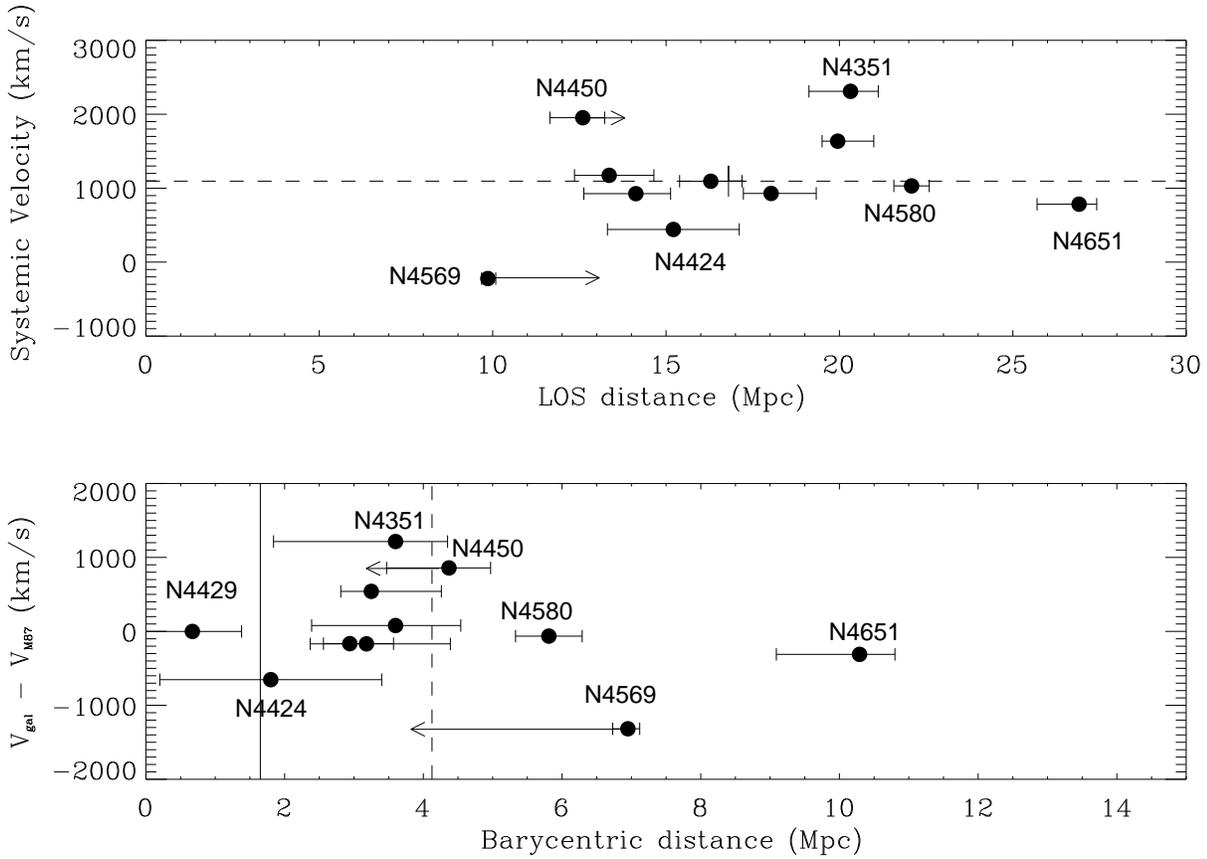}
%\plotone{fig10.eps}
}
\caption
{
Position-Velocity diagrams for sample galaxies in the Virgo cluster.
{\em Top panel:} LOS distances versus systemic velocities. Dashed line represents the systemic velocity of M87.
Cross represents the location of the core of the Virgo cluster (M87). {\em Bottom  panel:} 
Barycentric distances versus projected relative velocity with respect to M87. Solid line represents the virial radius, and
dash line represents the maximum rebound radius. Arrows represent
the expected location of NGC 4450 and NGC 4569 if a dark matter halo
is included. }\label{fig11}
\end{figure}

\begin{figure}
\figurenum{11}
\centerline{
\includegraphics[width=1.0\textwidth,clip]{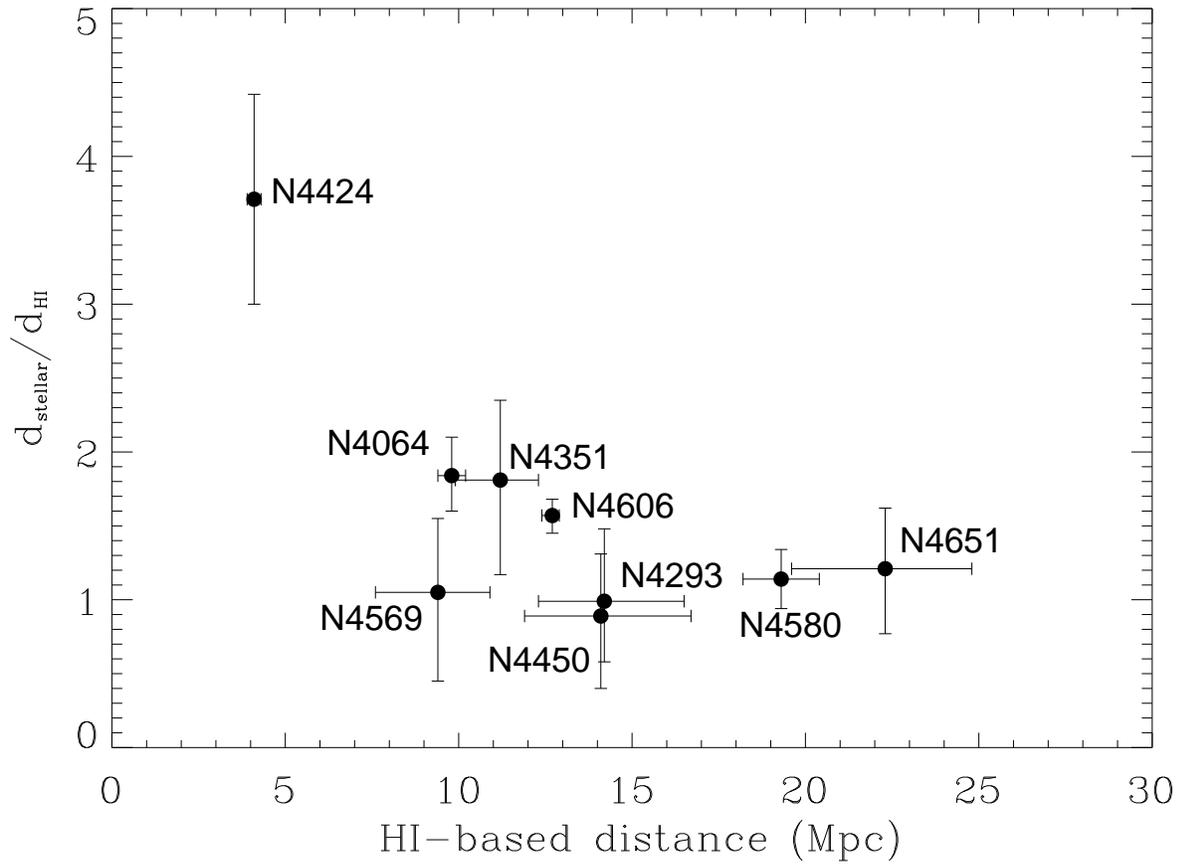}
}

\caption
{Ratio of SKB to HI based distances versus HI based distance.
Galaxies with small LOS HI based distances tend to have SKB distances at least 50\% higher. 
This shows that HI based distances can be biased by environmental effects.}\label{statdiff}
\end{figure}

NGC 4450, NGC 4569, NGC 4580, and NGC 4651 are galaxies further out from the
maximum rebound radius, according the SKB distances.
The first three galaxies are HI deficient.
NGC 4450 is an anemic galaxy with no evidence of recent ICM stripping. 
Its maximum circular velocity as
determined by us is lower than the maximum H$\alpha$ velocity, indicating that our
model contains less mass than needed for reproducing the observed H$\alpha$
velocities, probably due to either non-axisymmetric features which
were not included or a significant dark matter halo.
If we include a dark matter halo in order to match the H$\alpha$ rotation
curve (Fig $\ref{figdm}$), it is likely located inside
the maximum rebound radius ($d_{\rm bary} =$ 3.3 Mpc). 
NGC 4580 has a ``Truncated/Normal"
star formation class, so it is likely to have been affected by ICM--ISM
stripping.
However it couldn't have crossed the core according to the simple cluster model, and
the age of stellar population in outer disk reveals that it must have
been stripped outside the cluster core (Crowl \& Kenney 2008).
This galaxy is closer (in the plane of the sky) to M49 than M87, so perhaps it is
bound to the M49 subcluster rather than the M87 subcluster, and affected by the M49 environment.
The M49 subcluster is 4 times less massive than the M87 subcluster, and has a less dense ICM, 
so it is unlikely to strip the gas from a galaxy if the ICM is static and smooth, but could if
the ICM were dynamic and lumpy.
In general substructures in the cluster can be important in several ways: by
stirring up the ICM through sub-cluster mergers and locally increasing the strength of
ram pressure, by changing the value of the maximum rebound radius,
or by the pre-processing of galaxies in infalling groups. 

%%This is the only galaxy
%%in this sample HI-deficient and gas truncated surprisingly far from the core.
%% not true!  n4064 has same properties!

The case of NGC 4651 is clearer:  this galaxy is very far from the core of
Virgo (+10.3 $\pm_{1.2}^{0.5}$ Mpc), and its star formation class is ``Normal",
indicating that it
has not been affected by the ICM medium, so it probably correspond to a
background galaxy rather than a Virgo member. 

\newpage

Finally, the case of NGC 4569 is the most complicated. The SKB distance places
this galaxy 
outside of the maximum rebound radius ($r =$ -7.0 $\pm $0.2 Mpc), and
in front of the cluster. Its high HI deficiency, truncated HI and
H$\alpha$ distribution, and extraplanar gas arm (Vollmer etal 2004; Kenney etal 2008) are
clear evidence for recent ICM--ISM stripping. Simulations (Vollmer \etal 2004)
suggest that peak pressure occurred 300 Myrs ago. Its extreme line-of-sight
velocity (-1200 $\kms$ relative to the cluster mean) is consistent with a 
galaxy after peak pressure located on the
near side of the core, but it is hard to understand how it could
be as much as 6 Mpc closer than the core.
The prominent dust lanes are an issue here, biasing the distance determination, so it
might simply happen that this galaxy actually lies inside the maximum rebound
radius. Also, if we test the inclusion of a dark matter halo in order to match
the H$\alpha$ rotation curve (Fig $\ref{figdm}$), the corrected distance is about 13 Mpc,
putting this galaxy inside the maximum rebound radius.

\section{Conclusions}

In this paper, we have shown that stellar kinematics can be reliably used
for determining distances to spiral and peculiar galaxies. 
This technique presents a considerable
advantage with respect to gas kinematics based techniques since, unlike the latter, it is in
principle unaffected by physical processes other than gravity.

This technique can be considerably improved provided we are able
to overcome the biases described in \S 3.3:
\begin{itemize}
\item {\em Overcoming dust obscuration} by using near-infrared imaging.

\item {\em Extending observations of galaxies further out}. This
has the advantage of going beyond bars, and bulges, and getting a
better constraint on the dark matter halo.

\item {\em Improving the reliability of  synthetic rotation curves} by
increasing the number of sample galaxies used
to build these curves, and extend the sample to non-cluster undisturbed galaxies. 
\end{itemize} 

Using stellar kinematics techniques, we have obtained SKB distances to 11
peculiar Virgo Cluster galaxies. Our results can be summarized as follows:\\
\\
1.- Of the four galaxies in our sample with HI line-of-sight distances 
estimated by Solanes \etal (2002) to be significantly foreground to Virgo,
three galaxies have SKB distances within 4 Mpc of the cluster core,
only one (NGC 4569) has an SKB distance (within 3$\sigma$) consistent
with the HI LOS distance.\\
\\
2.- Galaxies with star-formation classes other than ``Truncated/Compact" have
SKB distances that agree with the HI-based
distances within
3$\sigma$ errors.
On the other hand, the ``Truncated/Compact" galaxies NGC 4064 and NGC 4424
exhibit large discrepancies
between HI-based and SKB distances. SKB predicts distances 2 to 4
times farther away than HI-based distances. Thus, environmental effects can
indeed affect both their gas content and kinematics, rendering HI-based
distances unreliable.\\
\\
3.- SKB distance determinations show that ``low rotator" galaxies
belong to the
cluster, strongly suggesting that environmental effects are the cause of the
their low
gas velocity. Galaxies with highly truncated gas disks and other peculiarities
such as disturbed stellar disks and non-circular or non-planar gas motions give
gas-based distances smaller than SKB distances. These galaxies
were perhaps subject to gravitational interactions plus ICM--ISM stripping.
On the other hand, galaxies with normal or truncated gas disks, and normal
stellar disks,
presumably affected only by ICM--ISM stripping, have SKB distances
consistent within the errors.with gas based distances.\\ 
\\
4.- Most of the galaxies in our sample are within the bounce--back zone
(e.g. distances within 4 Mpc of M87),
indicating they could have crossed the core of the cluster. If this is
the case,
ICM-ISM stripping could  explain the existence of HI-deficient galaxies
in the outskirts of the cluster, although there are some galaxies
that were clearly stripped outside of the core. This is probably due
to the existence of a dynamic lumpy ICM stirred up by sub-cluster
mergers, or gravitational interactions plus associated gas stripping.\\
\bigskip
\bigskip
\bigskip

We thank Roeland van der Marel for invaluable comments, crucial to this work and an anonymous referee for useful suggestions. Funding for this research has been provided
by Fundaci\'on Andes Chile, FONDAP project grant 15010003, Chile,
and NSF grant AST-0071251.

\onecolumn
\newpage

\clearpage

\clearpage

%\begin{figure}
%\figurenum{2}
%\psfig{bbllx=55pt,bblly=425pt,bburx=494pt,bbury=649pt,clip=,file=figures/mgeset2.ps}
%\caption{Continuated.
%From
%{\em left} to {\em right}; NGC 4651,
%and NGC 4694.}\label{fig5}
% \end{figure}
%\clearpage

\clearpage
%\begin{figure}
%\figurenum{3}
{\includegraphics[width=1.0\textwidth,clip]{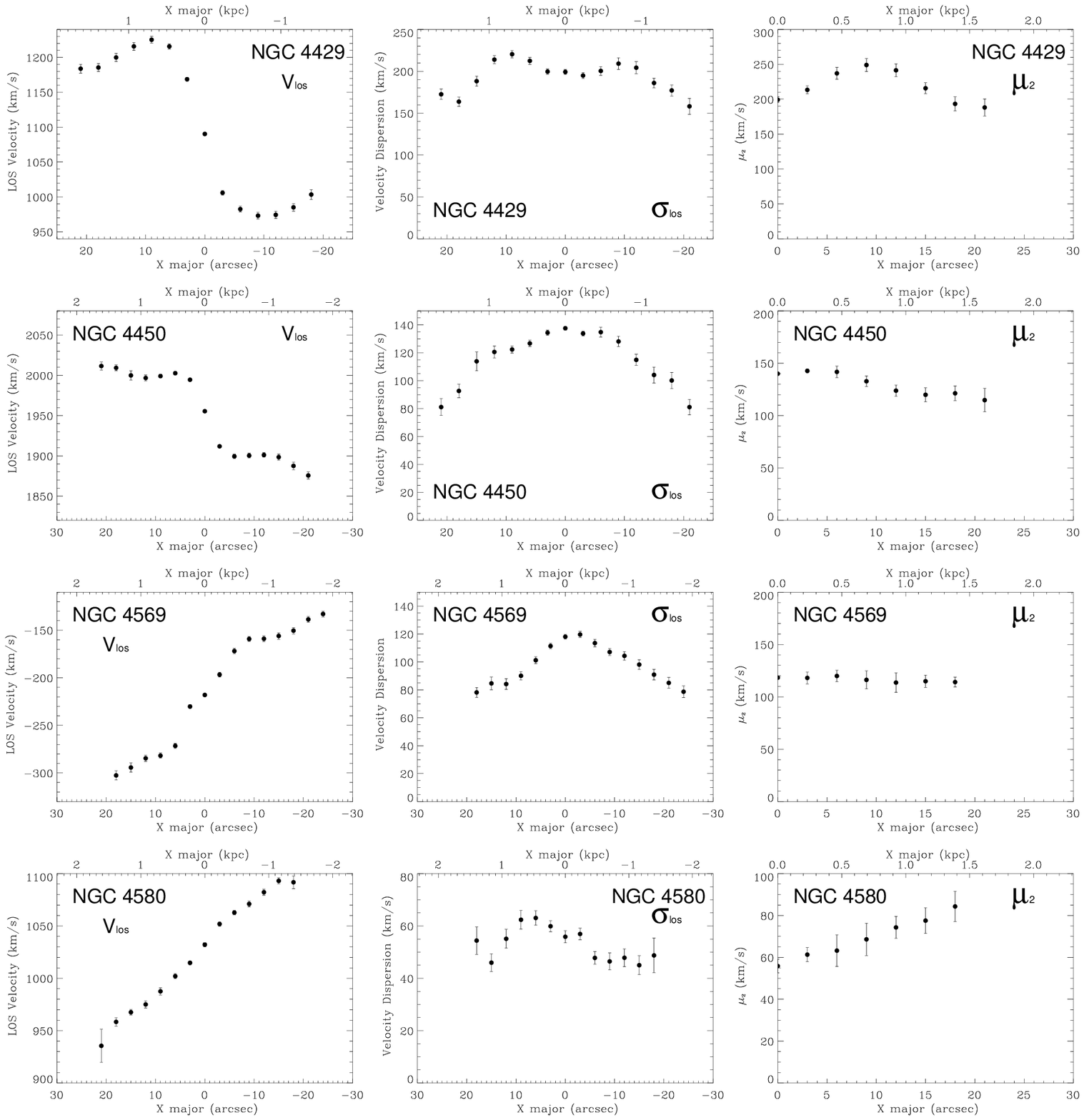}}\\
\centerline{Fig. 3. --- Continued.}
%\caption{Continued.
%From {\em top} to {\em bottom}; NGC 4429, NGC 4450, NGC 4569, and NGC 4580.}
\label{kindata2}
%\end{figure}
\clearpage
%\begin{figure}
%\figurenum{3}
{\includegraphics[width=1.0\textwidth,clip]{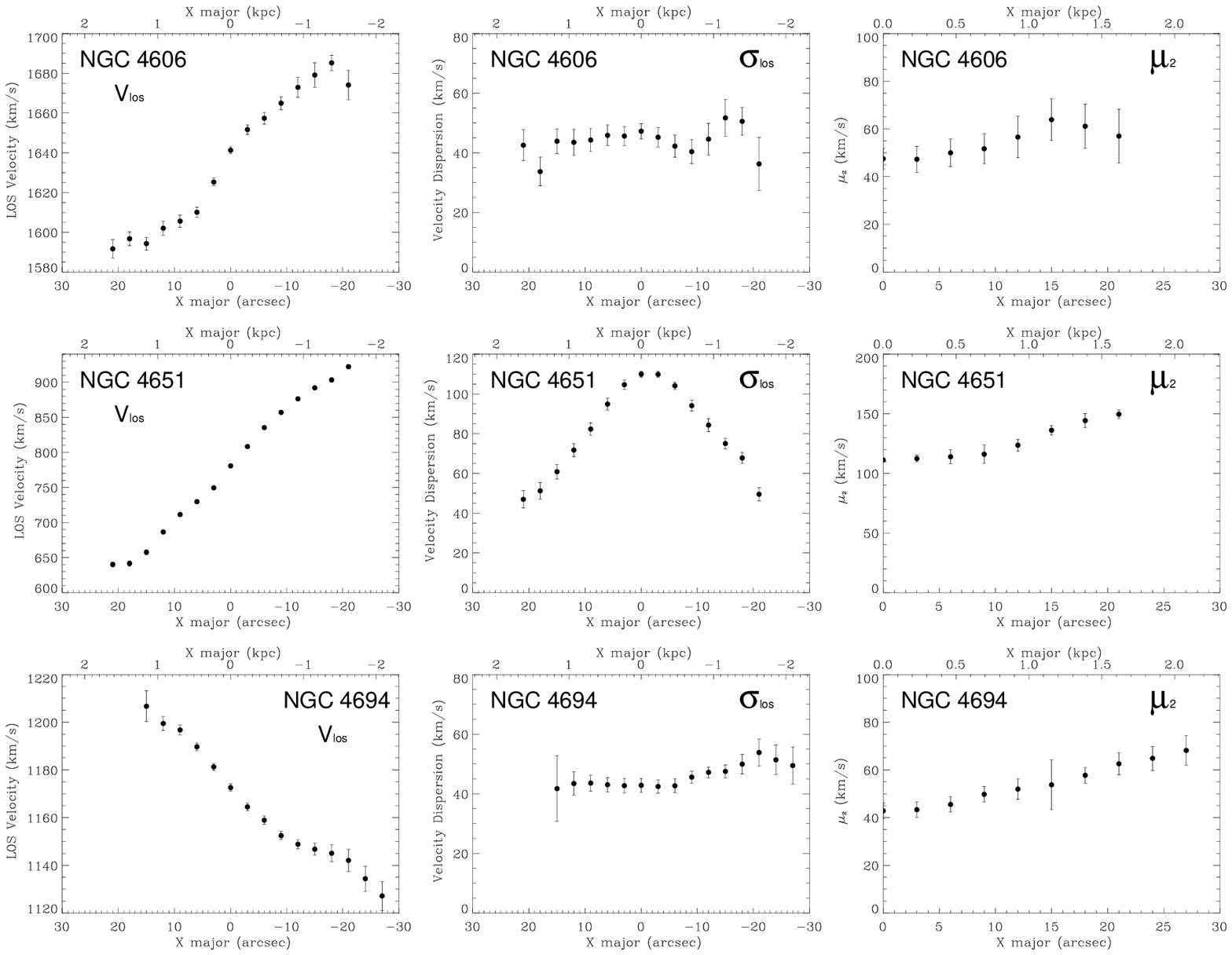}}\\
\centerline{Fig. 3. --- Continued.}
%\caption{Continued.
%From {\em top} to {\em bottom}; NGC 4606, NGC 4651, and NGC 4694.}
\label{kindata3}
%\end{figure}
\clearpage

\clearpage
{\plotone{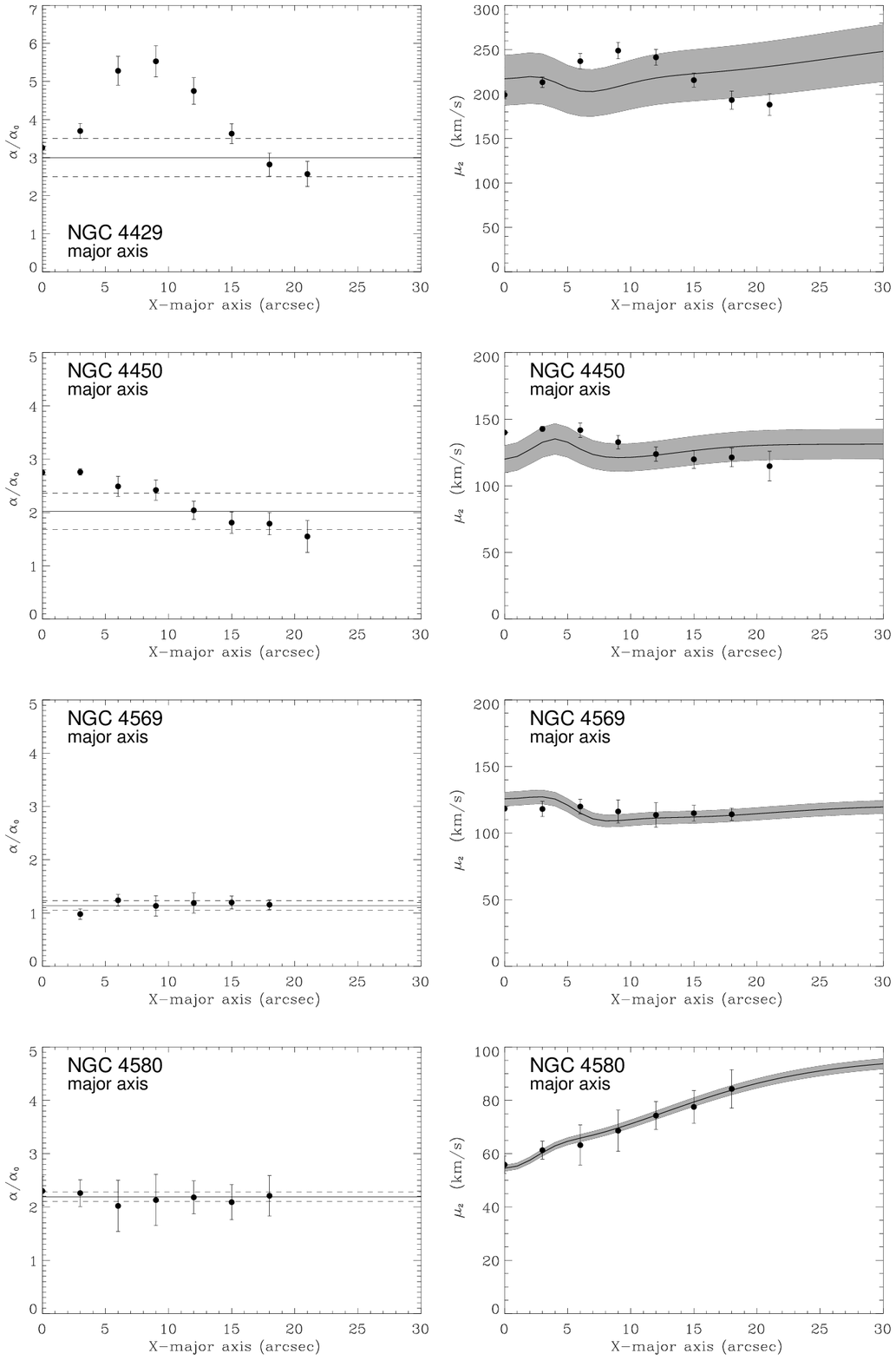}}\\
\centerline{Fig. 4. --- Continued.}
\label{fig7}
\clearpage
{\plotone{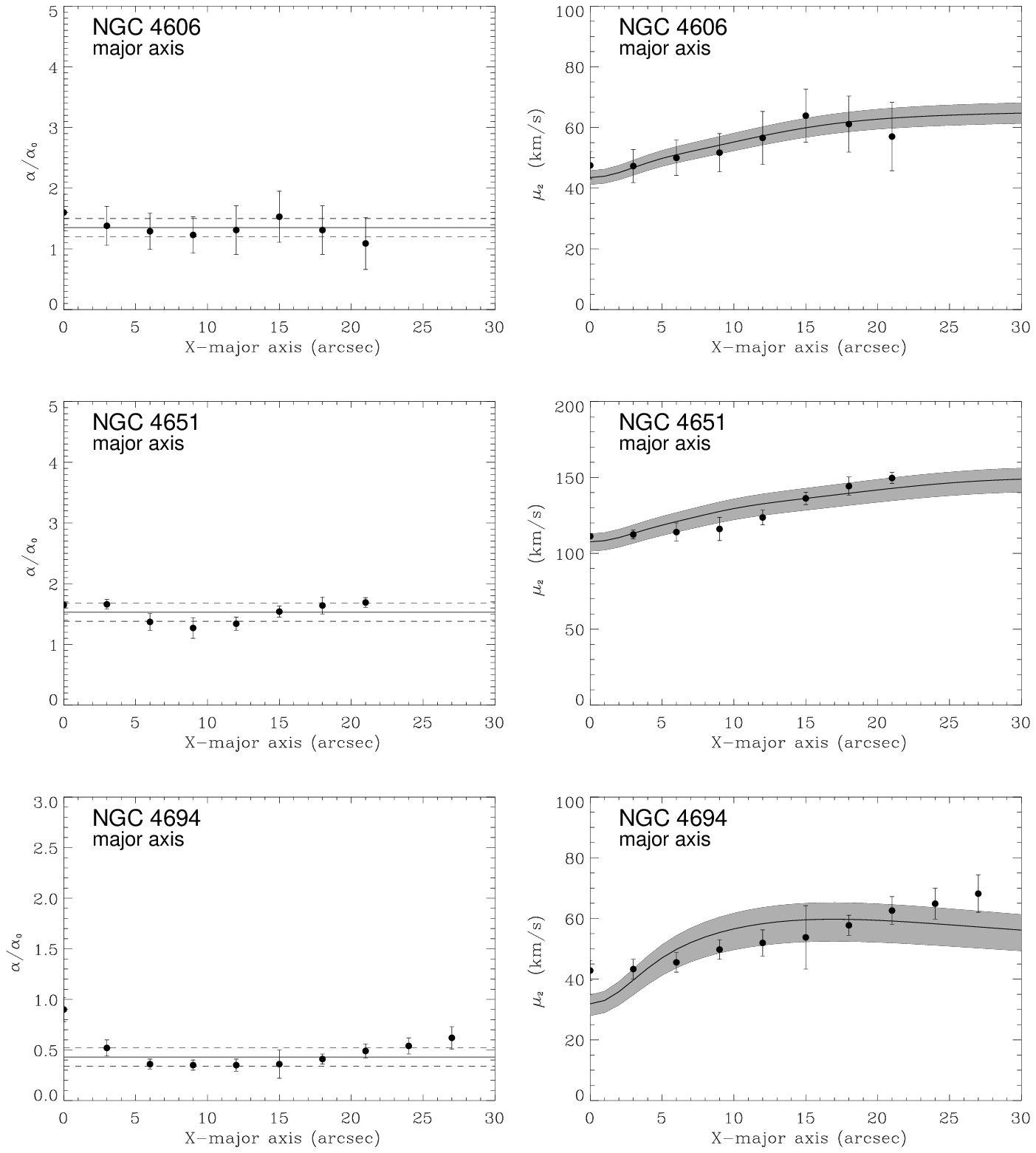}}\\
\centerline{Fig. 4. --- Continued.}
\label{fig8}
\clearpage

\clearpage
{\includegraphics[width=0.7\textwidth,clip]{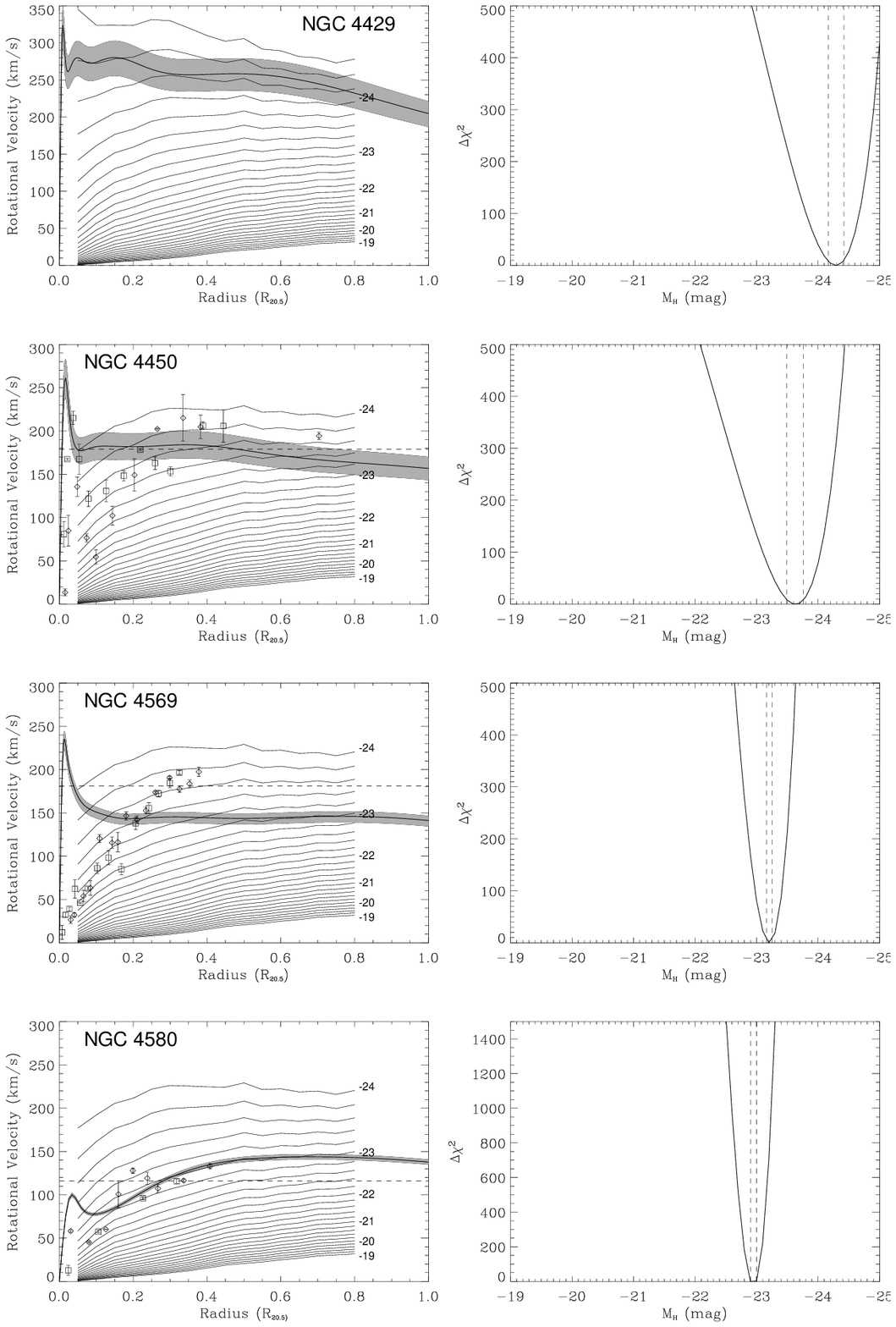}
}\\
\centerline{Fig. 7. --- Continued.}
\label{fig10}
\clearpage
{\includegraphics[width=0.7\textwidth,clip]{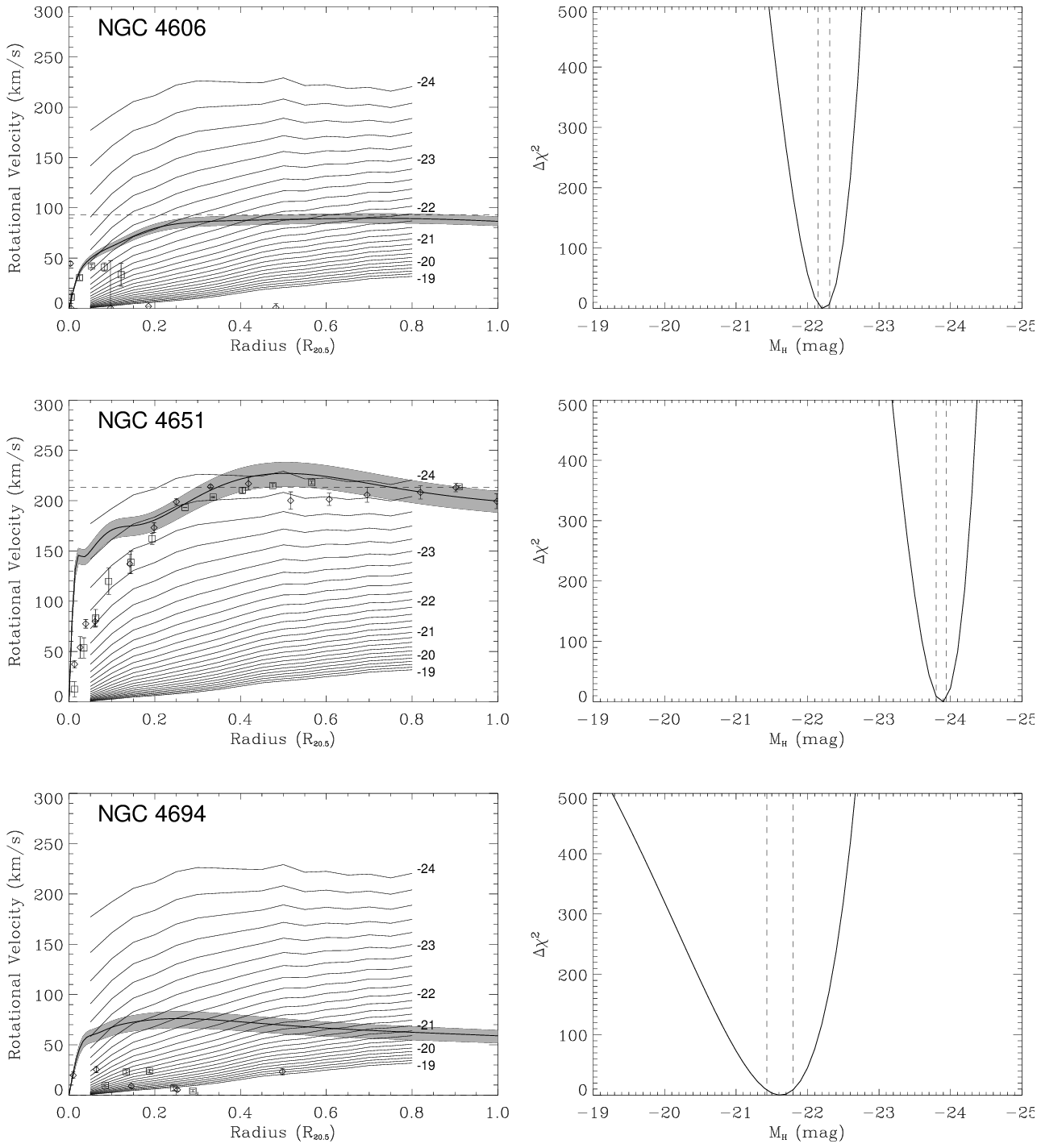}
}\\
\centerline{Fig. 7. --- Continued.}

\clearpage

\clearpage

\clearpage

\clearpage

\newpage

\begin{deluxetable}{ccccccccccc}
\tabletypesize{\scriptsize}
\tablecolumns{11}
%\tablenum{1}
\tablewidth{0pc}
\tablecaption{Galaxy Sample Properties \label{table1}}
\tablehead{
\colhead{}  &  \colhead{}  & \colhead{}  & \colhead{}  & \colhead{}  & \colhead{}  & \colhead{ }  & \colhead{Inc.} & \colhead{P.A} & \colhead{V$_{helio}$} & \colhead{D$_{M87}$} \\
\colhead{Name} & \colhead{R.A. (J2000)} & \colhead{Decl. (J200)} & \colhead{RSA/BST} & \colhead{RC3} & \colhead{SFC} & \colhead{B$_{T}^{0}$} & \colhead{(deg)} & \colhead{(deg)} & \colhead{($\kms$)} & \colhead{(deg)} \\
\colhead{(1)} & \colhead {(2)} & \colhead{(3)} & \colhead{(4)} & \colhead{(5)} & \colhead{(6)} & \colhead{(7)} & \colhead{(8)} & \colhead{(9)} & \colhead{(10)} & \colhead{11}}

\startdata
NGC 4064 & 12 04 11.2 &  18 26 36 & SBc(s): &    SB(s)a:pec  &  T/C  &  12.30 & 70 & 150 & 931 $\pm$  12 & 8.8 \\
NGC 4293 & 12 21 12.8 &  18 22 57  &   Sa pec  &   (R)SB(s)0/a &  T/A  &  11.20 & 67 & 66 & 930 $\pm$ 14 & 6.4\\
NGC 4351 & 12 24 01.6 &  12 12 18  &   Sc(s) II.3 &SB(rs)ab: pec &T/N[s] & 13.0 4& 47 & 80 & 2317 $\pm$ 16 & 1.7 \\
NGC 4424 & 12 27 11.5 &  09 25 15  &   Sa pec   &  SB(s)a:  &     T/C  & 12.32 & 60  & 90 & 440 $\pm$ 6 & 3.1 \\
NGC 4429 & 12 27 26.4 &  11 06 29  &   S0$_{3}$(6)/Sa pec &SA(r)0$^{+}$& \nodata & 10.9 & 62 & 90 & 1127 $\pm$ 31 & 1.5 \\
NGC 4450 & 12 28 29.3 &  17 05 07  &   Sab pec  &  SA(s)ab   &    T/A  & 10.93 & 46 & 175 & 1958 $\pm$ 6 & 4.7 \\
NGC 4569 & 12 36 49.8 &  13 09 46 &    Sab(s) I-II & SAB(rs)ab  &  T/N[s] & 10.2 5& 64 & 23 & -232 $\pm$ 22 & 1.7\\
NGC 4580 & 12 37 48.6 &  05 22 06  &   Sc/Sa   &    SAB(rs)a pec &T/N[s] & 12.49 & 45 & 158 & 1036 $\pm$ 7 & 7.2\\
NGC 4606 & 12 40 57.6 &  11 54 44  &   Sa pec   &   SB(s)a:  &    T/C & 12.69 & 67 & 40 & 1655 $\pm$ 16 & 2.5\\
NGC 4651 & 12 43 42.6 &  16 23 36  &   Sc(r) I-II & SA(rs)c  &     N   & 11.36&  51 & 71 & 804 $\pm$ 10 & 5.1 \\
NGC 4694 &  12 48 15.1 &  10 59 00  &   Amorph     & SB0 pec  &    T/N   & 12.19 & 42 & 140 & 1177 $\pm$ 11 & 4.5 \\
\enddata
\tablecomments{(1) Galaxy name; (2) Right Ascension in hours, minutes, and seconds; (3) Declination in degrees,
minutes, and seconds; (4) Hubble Types from BST, Sandage \& Tammann 1987, or Sandage \& Bedke 1994; (5)
Hubble type from RC3; (6) Star formation class from Koopmann \& Kenney 2004; (7) The total, face-on blue magnitude
from RC3; (8) Inclination from Koopmann \etal 2001; (9) Optical P.A from Koopmann \etal 2001; (10) Heliocentric
radial velocity from HyperLEDA; (11) The projected angular distance in degrees of thegalaxy from M87.}
\end{deluxetable}

\clearpage

\begin{deluxetable}{ccc}
%\tablecolumns{11}
%\tablenum{6.1}
\tablewidth{0pc}
\tablecaption{NIR Properties of Sample Galaxies \label{table4}}
\tablehead{
\colhead{}  & \colhead{$H_{c}$} & \colhead{$R_{20.5}$} \\
\colhead{Name} & \colhead{mag} & \colhead{arcsec}  \\
\colhead{(1)} & \colhead{(2)} & \colhead{(3)}}
\startdata
NGC 4064 & 8.78 & 64.0\tablenotemark{a}  \\
NGC 4293 & 7.35 & 156.30  \\
NGC 4351 & 10.44 & 35.50  \\
NGC 4424 & 9.17 & 75.69\tablenotemark{a} \\
NGC 4429 & 6.76 & 166.50 \\
NGC 4450 & 6.90 & 147.4 \\
NGC 4569 & 6.77 & 195.00 \\
NGC 4580 & 8.77 & 58.79 \\
NGC 4606 & 9.30 & 60.93 \\
NGC 4651 & 8.25 & 69.06 \\
NGC 4694 & 9.03 & 61.30 \\
\enddata
\tablenotetext{a}{2MASS Galaxy Atlas, Jarrett \etal 2003.}
\tablecomments{(1) Galaxy name; (2) Apparent magnitude in H-band (Gavazzi \etal 1999);
(3) Radius where the surface brightness is 20.5 mag arcsec$^{-2}$}
\end{deluxetable}

\clearpage

\begin{deluxetable}{cccccc}
\tablewidth{0pc}
\tablecaption{Best Model Parameters of Sample Galaxies \label{table5}}
\tablehead{
\colhead{} & \colhead{inc} & \colhead{} & \colhead{$\alpha_{0}$} & \colhead{$\alpha$} & \colhead{} \\
\colhead{Name} & \colhead{($\deg$)} & \colhead{$\alpha/\alpha_{0}$} & \colhead{$\Upsilon_{\odot}$ Mpc} & \colhead{$\Upsilon_{\odot}$ Mpc} & \colhead{Comments} \\
\colhead{(1)} & \colhead{(2)} & \colhead{(3)} & \colhead{(4)} & \colhead{(5)} & \colhead{(6)}}
\startdata
NGC 4064 & 70 & 1.82 $\pm$ 0.26 & 32 & 58 $\pm$ 8 & Major axis, mean $x \geq$ 5"\\
NGC 4293 & 70 &  2.68 $\pm$ 0.72 & 32 &   86 $\pm$ 23 & Major axis, mean $x \geq$ 5"\\
NGC 4351 & 47 &  1.12 $\pm$ 0.13 & 32 & 36 $\pm$ 4 & Major axis, mean $x \geq$ 5"\\
NGC 4424 & 72 & 1.17 $\pm$ 0.33 & 32 & 37 $\pm$ 11 & Major axis, mean $x \geq$ 5"\\
NGC 4429 & 66 & 3.0 $\pm$ 0.5 & 32 & 96 $\pm$ 16 & Major axis, mean $x \geq$ 12"\tablenotemark{a}\\
NGC 4450 & 50 & 2.02 $\pm$ 0.34 & 32 & 65 $\pm$ 1 & Major axis, mean $x \geq$ 5"\\ 
NGC 4569 & 66 & 1.14 $\pm$ 0.09 & 40 & 46 $\pm$ 4 & Major axis, mean $x \geq$ 5"\\
NGC 4580 & 46 & 2.19 $\pm$ 0.09 & 32 & 70 $\pm$ 3 & Major axis, mean $x \geq$ 5"
\\
NGC 4606 & 68 & 1.35 $\pm$ 0.15 & 32 & 43 $\pm$ 5 & Major axis, mean $x \geq$ 5"\\
NGC 4651 & 51 & 1.53 $\pm$ 0.15 & 55 & 84 $\pm$ 8 & Major axis, mean $x \geq$ 5"\\
NGC 4694 & 59 & 0.43 $\pm$ 0.09 & 32 & 14 $\pm$ 3 & Major axis, mean $x \geq$ 5"\\
\enddata
\tablenotetext{a}{Region dominated by circumnuclear disk was avoided}
\tablecomments{(1) Galaxy name; (2) Inclination used for obtaining oblate models; (3) Mean $\alpha/\alpha_{0}$ ratio; (4) Model $\alpha_{0}$; (5) Best $\alpha$ parameter; (6) Comments.}
\end{deluxetable}

\begin{deluxetable}{cccccccccc}
%\rotate
\tabletypesize{\scriptsize}
\tablewidth{0pc}

\tablecaption{Magnitudes and Distance Estimates\label{table6}}
\tablehead{
\colhead{} & \colhead{$M_{H}$} & \colhead{$\mu$} & \colhead{$d_{\rm stellar}$} & \colhead{$\Upsilon$} & \colhead{$\theta$} & \colhead{$d_{\rm bary}$} & \colhead{$d_{\rm HI}$} & \colhead{$V_{\rm sys}$} & \colhead{ }\\
\colhead{Name} & \colhead{mag} & \colhead{mag} & \colhead{Mpc} & \colhead{$\Upsilon_{\odot}$} & \colhead{($\deg$)} & \colhead{Mpc} & \colhead{Mpc} & \colhead{km s$^{-1}$} & \colhead{$d_{\rm stellar}/d_{\rm HI}$} \\
\colhead{(1)} & \colhead{(2)} & \colhead{(3)} & \colhead{(4)} & \colhead{(5)} & \colhead{(6)} & \colhead{(7)} & \colhead{(8)} & \colhead{(9)} & \colhead{(10)}}
\startdata
NGC 4064 & -22.5 $\pm_{0.2}^{0.1}$ & 31.3$\pm_{0.1}^{0.2}$ & 18.0 $\pm_{0.8}^{1.3}$ & 3.2 $\pm_{0.4}^{0.5}$ & 8.8 & 2.9 $\pm_{0.4}^{0.6}$ & 9.8 $\pm$ 0.4 & 929 $\pm$ 3 & 1.84 $\pm_{0.24}^{0.26}$ \\
NGC 4293 & -23.4 $\pm_{0.2}^{0.3}$ &30.8 $\pm_{0.3}^{0.2}$ & 14.1 $\pm_{1.5}^{1.0}$ & 6.1 $\pm_{1.8}^{1.7}$ & 6.4 & 3.2 $\pm_{0.8}^{1.2}$ & 14.2 $\pm_{1.9}^{2.3}$ & 926 $\pm$ 4 & 0.99 $\pm_{0.41}^{0.49}$ \\
NGC 4351 & -21.1 $\pm$ 0.1 & 31.5 $\pm$ 0.1 & 20.3 $\pm_{1.2}^{0.8}$ & 1.8 $\pm$ 0.2 & 1.7 & 3.6 $\pm_{1.8}^{0.8}$ & 11.2 $\pm_{1.3}^{1.1}$ & 2310 $\pm$ 2 & 1.8 $\pm_{0.6}^{0.5}$\\
NGC 4424 & -21.7 $\pm$ 0.3 & 30.9 $\pm$ 0.3 & 15.2 $\pm$ 1.9 & 2.4 $\pm$ 0.8 & 3.1 & 1.8 $\pm$ 1.6 & 4.1 $\pm$ 0.2 & 442 $\pm$ 4 & 3.7 $\pm$ 0.7\\
NGC 4429 & -24.3 $\pm$ 0.1 & 31.1 $\pm$ 0.1 & 16.3 $\pm$ 0.9 & 7.0 $\pm$ 1.9 &  1.5 & 0.7 $\pm$ 0.7 & \nodata & 1094 $\pm$ 6 & \nodata\\
NGC 4450\tablenotemark{a} & -23.6 $\pm_{0.2}^{0.1}$ & 30.5 $\pm_{0.1}^{0.2}$ & 12.6 $\pm_{0.6}^{1.0}$ & 5.9 $\pm$ 1.0 & 4.7 & 4.4 $\pm_{0.9}^{0.6}$ & 14.1 $\pm_{2.6}^{2.2}$ & 1953 $\pm$ 2 & 0.89 $\pm_{0.49}^{0.42}$\\
NGC 4569\tablenotemark{b} & -23.20 $\pm_{0.05}^{0.04}$ & 29.97 $\pm_{0.04}^{0.05}$ &9.9 $\pm$ 0.2 & 4.7 $\pm$ 0.4 & 1.7 & 7.0 $\pm$ 0.2 & 9.4 $\pm_{1.8}^{1.5}$ & -222 $\pm$ 6 & 1.1 $\pm_{0.6}^{0.5}$ \\
NGC 4580 & -22.95 $\pm$ 0.05 & 31.72 $\pm$ 0.05 & 22.1 $\pm$ 0.5 & 3.2 $\pm$ 0.2 & 7.2 & 5.8 $\pm$ 0.5 & 19.3 $\pm$ 1.1 & 1031 $\pm$ 4 & 1.1 $\pm$ 0.1 \\
NGC 4606 & -22.20 $\pm_{0.11}^{0.05}$ & 31.50 $\pm_{0.05}^{0.10}$ & 19.9 $\pm_{0.5}^{1.0}$ & 2.16 $\pm$ 0.3 & 2.5 & 3.25 $\pm_{0.4}^{1.0}$ & 12.7 $\pm_{0.3}^{0.2}$ & 1636 $\pm$ 5 & 1.6 $\pm$ 0.1 \\
NGC 4651 & -23.90 $\pm_{0.04}^{0.10}$ & 32.15 $\pm_{0.10}^{0.04}$ & 26.9 $\pm_{1.2}^{0.5}$ & 3.1 $\pm$ 0.3 & 5.1 & 10.3 $\pm_{1.2}^{0.5}$ & 22.3 $\pm_{2.7}^{2.5}$ & 784 $\pm$ 2 & 1.2 $\pm$ 0.4 \\
NGC 4694 & -21.6 $\pm$ 0.2 & 30.6 $\pm$ 0.2 & 13.4 $\pm_{1.0}^{1.3}$ & 1.1 $\pm$ 0.3 & 4.5 & 3.6 $\pm_{1.2}^{0.9}$ & \nodata & 1174 $\pm$ 22 & \nodata \\
\enddata
\tablenotetext{a}{If we add a dark matter halo in order to match the H$\alpha$ rotation curve we have that $M_{H} =$ -23.8 $\pm$ 0.1 mag, $\mu =$ 30.7 $\pm$ 0.1mag , and $d_{\rm stellar} =$ 13.8 $\pm_{0.6}^{0.8}$ Mpc }
\tablenotetext{b}{If we add a dark matter halo in order to match the H$\alpha$ rotation curve we have that $M_{H} =$ -23.8 $\pm$ 0.01 mag, $\mu =$ 30.57 $\pm$ 0.01 mag , and $d_{\rm stellar} =$ 13.0 $\pm$ 0.1 Mpc }
\tablecomments{(1) Galaxy name; (2) Absolute magnitude in H-band; (3) Distance modulus; (4) SKB distance; (5) Mass-luminosity ratio in the R-band; (6) Angular distance to M 87; (7) Barycentric SKB distance to M 87; (8) HI based distance; (9) Stellar systemic velocity; (10) Ratio between de SKB distance and HI based distance.}
\end{deluxetable}

\end{document}